\documentclass[sn-vancouver,Numbered]{sn-jnl}

\usepackage{graphicx}%
\usepackage{multirow}%
\usepackage{amsmath,amssymb,amsfonts}%
\usepackage{amsthm}%
\usepackage{mathrsfs}%
\usepackage[title]{appendix}%
\graphicspath{{./Figures/}}
\usepackage{subcaption}
\usepackage{xcolor}%
\usepackage{tabularx}
\usepackage{tabularray}
\usepackage{qcircuit}
\usepackage{textcomp}%
\usepackage{manyfoot}%
\usepackage{braket}
\usepackage{colortbl}
\usepackage{dcolumn}
\usepackage{booktabs}%
\usepackage{algorithm}%
\usepackage{algorithmicx}%
\usepackage{algpseudocode}%
\usepackage{listings}%

\begin{document}

\title[Article Title]{Quantum SMOTE with Angular Outliers: Redefining Minority Class Handling}

\author*[1]{\fnm{Nishikanta} \sur{Mohanty}}\email{nishikanta.m.mohanty@student.uts.edu.au}

\author[2]{\fnm{Bikash K.} \sur{Behera}}\email{bikas.riki@gmail.com}

\author[1]{\fnm{Christopher} \sur{Ferrie}}\email{Christopher.Ferrie@uts.edu.au}

\affil[1]{\orgdiv{Centre for Quantum Software and Information}, \orgname{University of Technology Sydney}, \orgaddress{\street{15 Broadway, Ultimo}, \city{Sydney}, \postcode{2007}, \state{NSW}, \country{Australia}}}

\affil[2]{\orgdiv{} \orgname{Bikash's Quantum (OPC) Pvt. Ltd.}, \orgaddress{\street{Balindi}, \city{Mohanpur}, \postcode{741246}, \state{WB}, \country{India}}}

\abstract{This paper introduces Quantum-SMOTEV2, an advanced variant of the Quantum-SMOTE method, leveraging quantum computing to address class imbalance in machine learning datasets without K-Means clustering. Quantum-SMOTEV2 synthesizes data samples using swap-tests and quantum rotation centered around a single data centroid, concentrating on the angular distribution of minority data points and the concept of angular outliers (AOL). Experimental results show significant enhancements in model performance metrics at moderate SMOTE levels (30–36\%), which previously required up to 50\% with the original method. Quantum-SMOTEV2 maintains essential features of its predecessor (\textit{arXiv:2402.17398}), such as rotation angle, minority percentage, and splitting factor, allowing for tailored adaptation to specific dataset needs. The method is scalable, utilizing compact swap tests and low-depth quantum circuits to accommodate a large number of features. Evaluation on the public Cell-to-Cell Telecom dataset with Random Forest (RF), K-Nearest Neighbours (KNN) Classifier, and Neural Network (NN) illustrates that integrating Angular Outliers modestly boosts classification metrics like accuracy, F1 Score, AUC-ROC, and AUC-PR across different proportions of synthetic data, highlighting the effectiveness of Quantum-SMOTEV2 in enhancing model performance for edge cases.}


\maketitle

\section{Introduction}\label{sec1}
Class imbalance poses a significant challenge in machine learning, especially when the distribution of classes within a dataset is skewed. This imbalance often results in models that favor the majority class, which can significantly impact the accurate prediction of the minority class \cite{haixiang_learning_2017, blaszczyk_framework_2021}. This problem is particularly prevalent in sectors like banking, insurance, and retail where fraud detection is critical, as well as in telecommunications for customer churn prediction and spam filtering in emails, where the class of interest is usually less represented. 

Among the various methods employed to counter this, the Synthetic Minority Oversampling Technique (SMOTE) \cite{wang_research_2021, chawla_smote_2002} emerges as one of the prominent algorithms. Originally introduced by Chawla et al. \cite{chawla_smote_2002}, SMOTE has been a cornerstone in addressing class imbalances. Our previous work advanced this approach by introducing Quantum-SMOTE \cite{Mohanty24_Q-SMOTE}, which adapts SMOTE to quantum computing, moving away from traditional methods like KNN \cite{Cover_KNN_1967} and Euclidean distances for generating synthetic samples.

This paper introduces a refined variant of Quantum-SMOTE \cite{Mohanty24_Q-SMOTE} that optimizes the original method and brings forth the novel concept of angular distribution and Angular Outliers (AOL). For general clarity, we name this new variant of Quantum-SMOTE as Quantum-SMOTEV2. 

Unlike conventional machine learning and statistical methods that focus on analyzing individual feature distributions, our method considers the overall spatial distribution of data points within the feature space. We suggest that examining the angular distribution relative to the data centroid can offer a comprehensive view of data point distributions across all features. This perspective enables the detection of outlier patterns, which could represent critical edge cases, thereby enhancing the robustness of classification models. The result is a noticeable improvement of classification effectiveness over moderate levels of SMOTE as we gradually test the effectiveness of Angular Outlier boosting with an increase in the percentage of synthetic samples. In our experiment we are able to see significant improvements of model performance parameters at moderate levels of smote around (30–36\%) with AOL, what was originally possible with full SMOTE at 50\%.
We tested this improved methodology on a different dataset, the cell-to-cell churn dataset \cite{cell2cell_telco_churn_dataset_kaggle}, and assessed its performance using three well-known classification algorithms: Random Forest (RF), K-Nearest Neighbours (KNN), and Neural Networks (NN). The combination of RF, KNN, and NN provides a balanced representation of three fundamentally different algorithmic paradigms. Ensemble-Based RF tests how boosting techniques integrate with models that rely on aggregated decision-making. Instance-Based KNN evaluates the impact of Quantum-SMOTE and AOL on algorithms highly sensitive to data distribution and neighborhood structure, Model-Based NN explores how synthetic data and outlier adjustments enhance complex, non-linear decision-making processes. The dataset \cite{cell2cell_telco_churn_dataset_kaggle} was specifically chosen due to its propensity to produce biased models if not properly balanced, thus making it ideal for testing.

The paper is structured in the following manner: Section \ref{Background} explores the core principles of the Quantum-SMOTEV2 algorithm and the concept of angular outliers, along with an overview of model evaluation metrics. Section \ref{Methodology} provides an analysis of the creation of Quantum-SMOTEV2 via the use of quantum swap test and quantum rotation concepts. The section also discusses the process of finding angular outliers and boosting them to improve existing Quantum-SMOTE. Section \ref{Case Study and Results} relates to the implementation of the Quantum-SMOTEV2 algorithm on the cell-to-cell churn dataset \cite{cell2cell_telco_churn_dataset_kaggle}. This process comprises data preparation, the production of synthetic data using the Quantum-SMOTE method, and the boosting of Angular Outliers. We utilize the SMOTE with angular outliers technique on the cell-to-cell data, varying the proportions of the minority class from 30-50\%, accessing the impact of smote on various models and corresponding changes when outlier boosting is employed. In Section \ref{Inferences}, we summarize the results and model parameters of the classification models, which elucidate the effects of Quantum-SMOTEV2 and Angular Outliers.

\section{Background} \label{Background}

In our prior work, we introduced Quantum-SMOTE \cite{Mohanty24_Q-SMOTE}, a novel approach that, while fundamentally different in its mechanics from the traditional SMOTE \cite{chawla_smote_2002}, serves the same purpose of addressing class imbalances. Quantum-SMOTE synthesizes new data points by determining the angle between a minority data point and the data centroid(cluster centroid) and then adjusts this angle using a randomly assigned weight before rotating the original point to generate a new, synthetic one. 

This method employs quantum processes such as Quantum-SWAP tests and Quantum-rotational circuits, ensuring that rotation angles remain minimal to prevent the synthetic samples from straying too far from their original points. Consequently, these synthetic samples enhance the density of minority class points in specific data regions rather than merely positioning new points linearly between two neighboring minority points, a method used by classical SMOTE that relies on KNN and Euclidean distance. These generated samples thereby help increase the minority class’s representation, effectively mitigating bias towards the majority class in classification tasks. The figures \ref{QSMOTE_Mechanisms} illustrate the mechanism of Quantum SMOTE. 

In this paper, we propose Qunatum-SMOTEV2 which retains all the features of the previous version but removes the essential pre-step of clustering, thereby relying on a single data centroid to generate synthetic data samples. 

The proposed Quantum-SMOTEV2 eliminates clustering, hence eliminating the need for multiple centroids to produce synthetic samples; instead, this may be accomplished using a single data centroid for the whole dataset. This process calculates the angles between the dataset centroid and minority data points, allowing for the reliable observation that certain minority data points are closer to the centroid while others are farther away. 

It is conceivable that the minority data points located farther from the centroid are poorly distributed. By analyzing the distribution of these angles, we can discern the distribution features and find the outlier data points that fall beyond the interquartile range (IQR). We designate the data points that go beyond the upper/lower bound $\pm$ interquartile range (IQR) by 1.5 as outliers, referring to them as Angular Outliers. This research will evaluate the effect of enhancing angular outliers on classification performance across varied proportions of SMOTE.

Below, we review the figures of merit discussed in the evaluation of our proposal. 

\subsection{Model Evaluation Metrics}
\subsubsection{Confusion Matrix}
The confusion matrix is a tabular representation that encapsulates the efficacy of a classification model by displaying the frequencies of various prediction kinds. It serves as the foundation for several indicators used in ROC analysis.
\begin{itemize}
    \item True Positives (TP): Accurately identified positive instances.
    \item False Negatives (FN): Positive instances that were erroneously labelled as negative.
    \item False Positives (FP): Erroneously identified positive instances that were, in fact, negative.
    \item True Negatives (TN): Accurately identified negative instances.
\end{itemize}

Essential metrics Extracted from the Confusion Matrix:

\textbf{Accuracy:} Assesses the ratio of right predictions (including true positives and true negatives) to the total number of forecasts made.
\begin{equation}
  Accuracy = \frac{TP + TN}{TP + TN + FP + FN} 
\end{equation}

\textbf{Precision}: Concentrates on the proportion of accurately anticipated positive instances among all expected positive cases.
\begin{equation}
Precision = \frac{TP}{TP + FP}
\end{equation}

\textbf{Recall (TPR/Sensitivity)}: Assesses the model's efficacy in identifying true positives, as previously stated.
\begin{equation}
  TPR = \frac{TP}{TP + FN}  
\end{equation}

\textbf{F1 Score:} The harmonic mean of accuracy and recall, advantageous for achieving equilibrium between these two measurements.
\begin{equation}
 F1 = 2  \times(Precision \times Recall)   
\end{equation}

\subsubsection{ROC}
The Receiver Operating Characteristic (ROC) curve is a reliable tool for assessing the performance of a binary classification model. The model's performance variation is shown when the decision threshold is adjusted. The objective is often to achieve an optimal balance between recognising true positives (accurate predictions) and reducing false positives (incorrect predictions).

\textbf{True Positive Rate (TPR)}, sometimes referred to as Recall or Sensitivity, quantifies the number of real positive instances accurately detected by the model.

\textbf{False Positive Rate (FPR)}: This quantifies the frequency at which the model erroneously identifies a positive instance while the true class is negative.
\begin{eqnarray}
FPR = \frac{FP}{FP + TN}
\end{eqnarray}
The ROC curve illustrates the True Positive Rate (TPR) on the y-axis vs the False Positive Rate (FPR) on the x-axis for various threshold levels. As the threshold varies, the trade-off between the two measurements becomes evident.
By comprehending the confusion matrix and ROC curve in conjunction, one may more effectively assess the merits and shortcomings of the model. A model with high Recall but poor Precision may identify the majority of positive instances while also generating many false positives (high FPR).
\subsubsection{AUC}
The Area Under the Curve (AUC) quantifies the model's overall performance. AUC values approaching 1 indicate the model's proficiency in differentiating between the two classes, whilst values around 0.5 suggest performance equivalent to random chance.

\section{Emulating Quantum-SMOTEV2 with angular outliers } \label{Methodology}
In our previous research \cite{Mohanty24_Q-SMOTE}, we presented a comprehensive methodology for generating synthetic data for minority classes using quantum processes. The approach involved dynamically segmenting the entire population through clustering techniques and generating synthetic data within each cluster to achieve the desired minority class ratio. Specifically, quantum processes such as the SWAP test and controlled rotation were used to create synthetic data, with clustering serving as a critical step for dynamic segmentation.
Although this method effectively addresses class imbalance across various classification algorithms, we observed that using controlled rotations with small angles may benefit from an alternative approach. Instead of multiple cluster centroids, applying the method with a single data centroid could simplify the process by eliminating the need for initial clustering.

When generating synthetic samples around a single data centroid, we observed an interesting phenomenon: some data points were positioned closer to the centroid, while others were farther away. This variation created a distinct distribution of data points based on their angular distance from the centroid. Notably, the angle formed between a data point and the data centroid emerges as a comprehensive representation of the data point within the feature space, effectively capturing all its features in a unified manner. Traditionally, in machine learning, individual features are treated as having their own distributions and characteristics, but no single feature can holistically represent the distribution of a data point. In contrast, the angular distance offers a multidimensional perspective, encapsulating the contributions of all features to describe the data point in feature space.

This angular distance distribution reveals valuable insights into the structure of minority class data. Specifically, our analysis focuses on the outliers within this angular distribution, as they demonstrate a notable impact on model performance. As elaborated on above, we define these outliers as data points whose angular distances exceed 1.5 times the interquartile range (IQR) beyond (upper(75\%)/lower(25\%) bound) of the angular distance distribution.

In this study, we propose a method to enhance the impact of these angular outliers after generating synthetic data. The motivation behind this approach lies in the fact that outliers, as sparsely located data points, are often ignored by the decision boundary, leading to potential false positives in model evaluation. Since minority populations in many industrial contexts are inherently sparse, reducing false positives can play a crucial role in improving the reliability of predictive models.
With this new approach, we retain all the features and advantages of Quantum-SMOTE, such as rotation angle, minority percentage, and splitting factor, and also introduce new parameters for angular outlier boosting.

\begin{figure}[H]
\centering
\begin{subfigure}{0.8\linewidth}
\includegraphics[width=\linewidth]{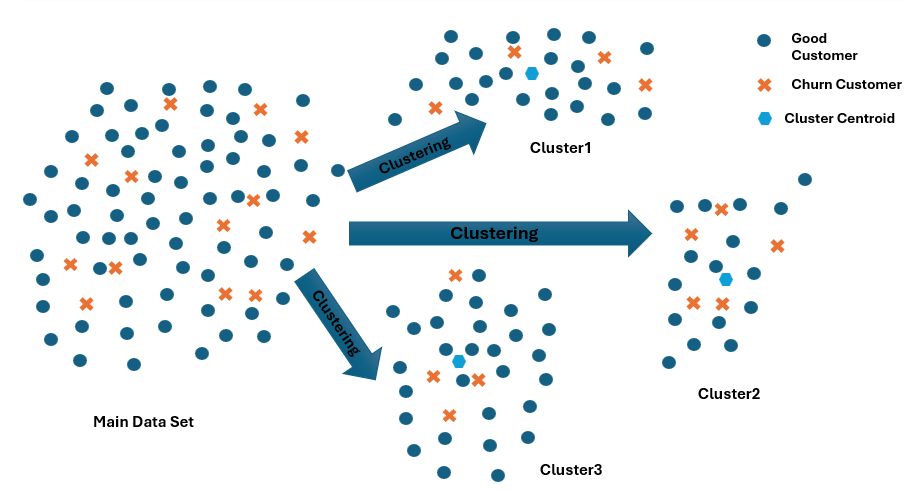} 
\caption{}
\label{classical_SMOTE}
\end{subfigure}\hfill
\begin{subfigure}{0.8\linewidth}
\includegraphics[width=\linewidth]{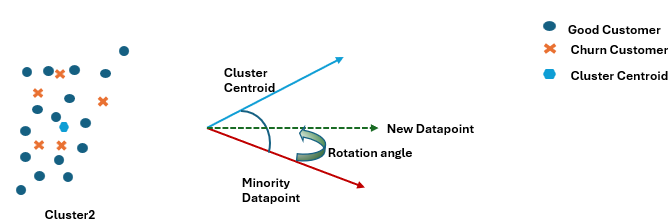} 
\caption{}
\label{Quantum_SMOTE}
\end{subfigure}\hfill
\caption{Plot illustrating different SMOTE mechanisms. (a) Data Clustering, (b) Quantum SMOTE.}
\label{QSMOTE_Mechanisms}
\end{figure}

\begin{figure} [H]
    \centering
    \includegraphics[width=0.7\linewidth]{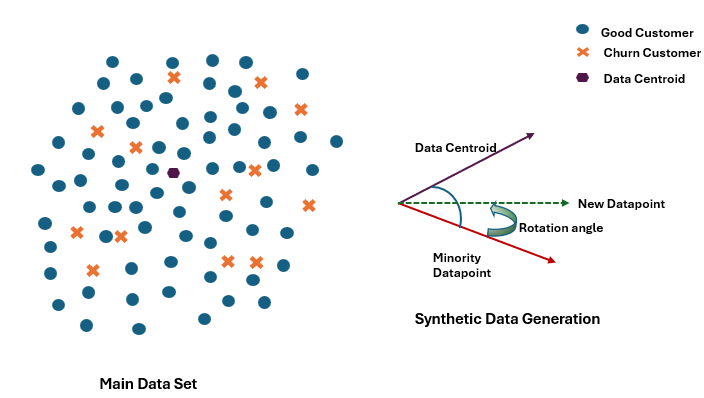}
    \caption{Proposed Quantum-SMOTEV2  with a single data centroid.}
    \label{QSMOTE2_Single_Centroid}
\end{figure}

Figures \Ref{QSMOTE_Mechanisms} and \ref{QSMOTE2_Single_Centroid} illustrate fundamental difference in Quantum Smote and proposed variant Quantum-SMOTEV2 .

\subsection{Compact Swaptest} \label{Compact Swaptest}

The quantum swap test is a method used in quantum computing to determine how similar two quantum states, $\psi$ and $\phi$, are. The outcome of the test reflects the extent of overlap between these two states, which is mathematically represented by their inner product, $\braket{\psi|\phi}$.

In this research, as well as in our previous work, we employ a modified version of the swap test to calculate the inner product between two vectors: the data centroid and a selected minority data point from the dataset. This process is outlined in detail in prior works, such as \cite{qiskit_medium, Mart_Dissimilarity_2023,Mohanty24_Q-SMOTE}. Although the referenced articles describe the method as a measure of dissimilarity and use it to compute Euclidean distance, we have adapted it to calculate the inner product of quantum states, which in turn helps us measure angular distance.

One of the advantages of this method is that it requires fewer qubits, specifically $$n=\log _2(M)+1$$
where n is the number of qubits and M is the classical data encoded using amplitude embedding. The procedure follows the steps outlined below.

We amplitude encode two vectors DC (Centroid) and MD (Minority) by 

\begin{eqnarray}
DC \longrightarrow|DC\rangle & =&\frac{1}{|DC|} \sum_i DC_i\left|q_i\right\rangle \\
MD \longrightarrow|MD\rangle & =&\frac{1}{|MD|} \sum_i MD_i\left|q_i\right\rangle
\end{eqnarray}

We define the quantum states $|\psi\rangle$ and $|\phi\rangle$ as:

\begin{eqnarray}
|\psi\rangle&=&\frac{|0\rangle \otimes|DC\rangle+|1\rangle \otimes|MD\rangle}{\sqrt{2}}  \nonumber \\
|\phi\rangle&=&\frac{|DC||0\rangle-|MD||1\rangle}{\sqrt{Z}} \nonumber \\
Z&=&|DC|^2 + |MD|^2
\end{eqnarray}

Let us calculate inner product of $\psi$ and $\phi$,

\begin{eqnarray}
\langle\phi \mid \psi\rangle=\left(\frac{\langle DC| \otimes\langle 0|-\langle MD| \otimes\langle 1|}{\sqrt{Z}}\right)\left(\frac{|0\rangle \otimes|DC\rangle+|1\rangle \otimes|MD\rangle}{\sqrt{2}}\right)
\end{eqnarray}

Expanding the inner product:
\begin{eqnarray}
\langle\phi \mid \psi\rangle&=&\frac{1}{\sqrt{Z}} \frac{1}{\sqrt{2}}(\langle DC| \otimes\langle 0|(|0\rangle \otimes|C\rangle)+\langle DC| \otimes\langle 0|(|1\rangle \otimes|MD\rangle)\nonumber\\
&-&\langle MD| \otimes\langle 1|(|0\rangle \otimes|DC\rangle)-\langle MD| \otimes\langle 1|(|1\rangle \otimes|MD\rangle))
\end{eqnarray}

Simplifying each term:
\begin{eqnarray}
&&\langle DC| \otimes\left\langle\left. 0|(|0\rangle \otimes|DC\rangle)=\langle DC \mid DC\rangle \otimes\langle 0 \mid 0\rangle=| C\right|^2\right. \nonumber \\
&&\langle DC| \otimes\langle 0|(|1\rangle \otimes|MD\rangle)=0  \nonumber \\ 
&&\langle MD| \otimes\langle 1|(|0\rangle \otimes|CD\rangle)=0 \nonumber \\
&&\langle MD| \otimes\left\langle\left. 1|(|1\rangle \otimes|MD\rangle)=\langle MD \mid M\rangle \otimes\langle 1 \mid 1\rangle=| MD\right|^2\right.
\end{eqnarray}

So, the inner product simplifies to:
\begin{eqnarray}
\langle\phi \mid \psi\rangle&=&\frac{1}{\sqrt{Z}} \frac{1}{\sqrt{2}}\left(|DC|^2-|MD|^2\right) \nonumber \\
\langle\phi \mid \psi\rangle&=&\frac{|DC|^2-|MD|^2}{\sqrt{2 Z}}
\end{eqnarray}
Calculating $|\langle\phi \mid \psi\rangle|^2$ :
\begin{eqnarray}
|\langle\phi \mid \psi\rangle|^2=\left(\frac{|DC|^2-|DM|^2}{\sqrt{2 Z}}\right)^2=\frac{\left(|DC|^2-|MD|^2\right)^2}{2 Z}
\end{eqnarray}

\begin{eqnarray}
{2 Z}|\langle\phi \mid \psi\rangle|^2=2 Z\left(\frac{\left(|DC|^2-|M|^2\right)^2}{2 Z}\right)
\end{eqnarray}

simplifying:
\begin{eqnarray}
{2 Z}|\langle\phi \mid \psi\rangle|^2=\left(|DC|^2-|MD|^2\right)^2
\end{eqnarray}

Assuming 
\begin{eqnarray}
{2 Z}|\langle\phi \mid \psi\rangle|^2 &=& D^2\nonumber\\
\implies D^2&=&2 Z|\langle\phi \mid \psi\rangle|^2
\end{eqnarray}
The term $D$ represents the Euclidean distance \cite{Mart_Dissimilarity_2023}, and the inner product of $\langle\phi \mid \psi\rangle$ represents the swaptest probability.

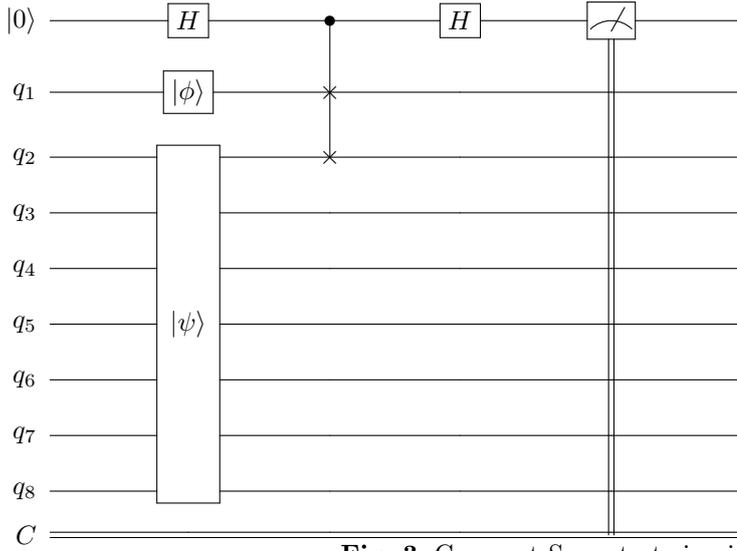
\begin{figure}[H]
\centering
\Qcircuit @C=4em @R=1.2em {
  \lstick{\ket{0}} & \gate{H} & \ctrl{2} & \gate{H} & \meter & \qw \\
  \lstick{q_1} & \gate{\ket{\phi}} & \qswap & \qw & \qw & \qw \\
  \lstick{q_2} & \multigate{6}{\ket{\psi}} & \qswap & \qw & \qw & \qw \\
  \lstick{q_3} & \ghost{\ket{\psi}} & \qw & \qw & \qw & \qw \\
  \lstick{q_4} & \ghost{\ket{\psi}} & \qw & \qw & \qw & \qw \\
  \lstick{q_5} & \ghost{\ket{\psi}} & \qw & \qw & \qw & \qw \\
  \lstick{q_6} & \ghost{\ket{\psi}} & \qw & \qw & \qw & \qw \\
  \lstick{q_7} & \ghost{\ket{\psi}} & \qw & \qw & \qw & \qw \\
  \lstick{q_8} & \ghost{\ket{\psi}} & \qw & \qw & \qw & \qw \\
  \lstick{C} & \cw & \cw & \cw & \cw \cwx[-9] & \cw 
}
\caption{Compact Swap test circuit.}
\label{Compact Swap test circuit.}
\end{figure}

In light of this, we define the angular distance—the angle between two vectors—as follows:
\begin{eqnarray}
\text{angular\_distance} = 2 \cos^{-1}(\sqrt{\text{swap\_test\_probability}})
\end{eqnarray}
This angular distance, or the angle between the two vectors, will be used to rotate the minority class data point, as we will explain in the following sections.

\subsection{Rotation of Minority data point} \label{Methodology Rotation}
Upon determining the angle (angular distance) between two vectors, we rotate the actual minority data point by an angle less than the predicted angle to generate a synthetic data point. We choose to reduce the degree of rotation to avert sudden variations in the minority data point values. In our previous work, we accessed the rotation of minority data points in X, Y and Z rotations and discussed their impacts \cite{Mohanty24_Q-SMOTE}. Thus we are not covering the details in this paper rather we just present the procedure below

\begin{algorithm} 
\caption{Angle of rotation calculation logic \cite{Mohanty24_Q-SMOTE}}
\label{rotation_logic}
\begin{algorithmic}[H]
\State \textbf{sf}: split\_factor
\If{$\text{angular\_distance} > \frac{\pi}{2}$}
    \State $\text{angle} \gets \left| \frac{\pi}{2} - \text{angular\_distance} \right| / \text{sf}$
\ElsIf{$\text{angular\_distance} < 0$}
    \State $\text{angle} \gets \left| \left( \frac{\pi}{2} - \text{angular\_distance} \right) \times \text{random}(0.5, 1) \right| / \text{sf}$
\Else
    \State $\text{angle} \gets \text{random}(0, \text{angular\_distance}) / \text{sf}$
\EndIf
\end{algorithmic}
\end{algorithm}

\begin{figure}[H]
\centering
\begin{subfigure}{\linewidth}
\includegraphics[width=0.8\linewidth]{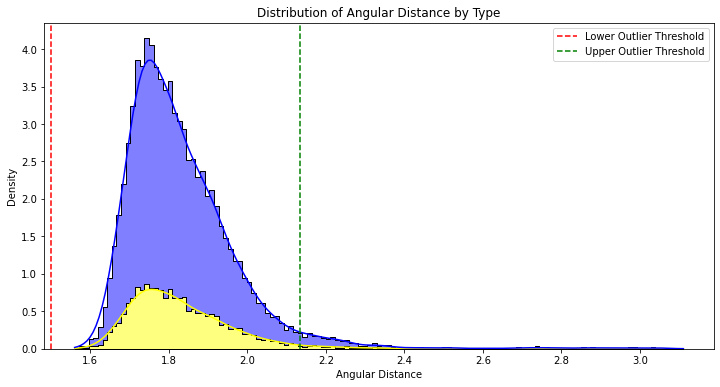} 
\caption{}
\label{Angular_Distribution}
\end{subfigure}\hfill
\begin{subfigure}{\linewidth}
\includegraphics[width=0.8\linewidth]{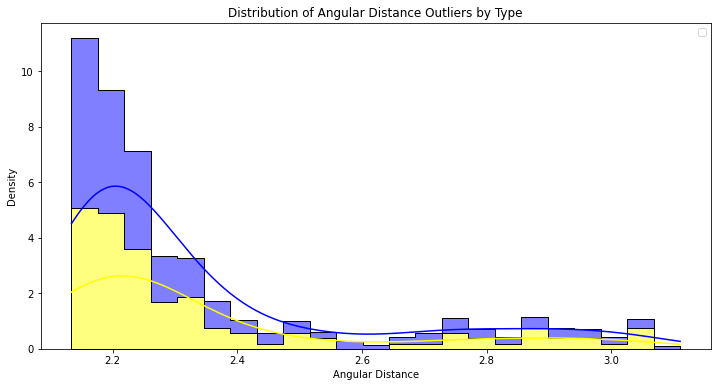} 
\caption{}
\label{Outlier_Distribution_2d}
\end{subfigure}\hfill
\caption{Figure showing idea of angular distribution with blue region showing original data and yellow region being synthetic data (a) Angular Distribution and Outlier Regions. (b) Distribution of Outliers in one of Outlier regions .}
\label{qsmote_distribution_outliers}
\end{figure}

\begin{figure}[H]
\centering
\begin{subfigure}{0.5\linewidth}
\includegraphics[width=\linewidth]{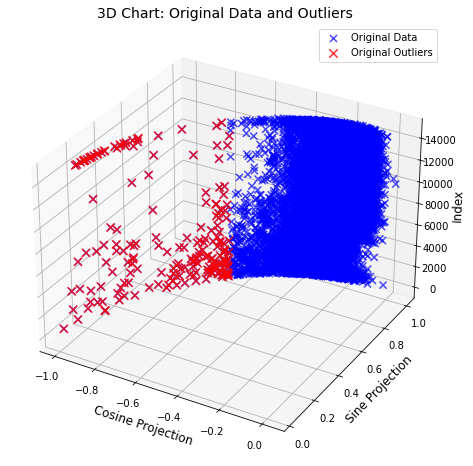} 
\caption{}
\label{Original_outlier}
\end{subfigure}\hfill
\begin{subfigure}{0.5\linewidth}
\includegraphics[width=\linewidth]{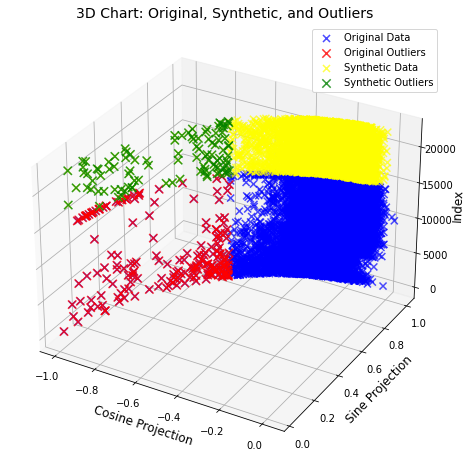} 
\caption{}
\label{Original_synthtic_outlier}
\end{subfigure}\hfill
\begin{subfigure}{0.5\linewidth}
\includegraphics[width=\linewidth]{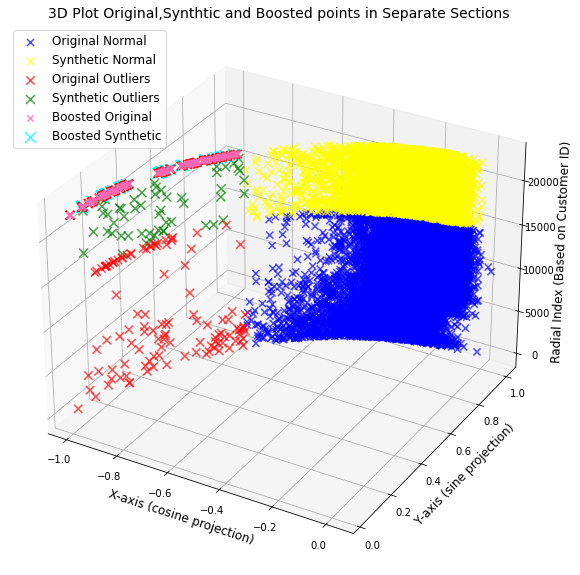} 
\caption{}
\label{Original_synthtic_boosted_outlier}
\end{subfigure}\hfill
\begin{subfigure}{0.5\linewidth}
\includegraphics[width=\linewidth]{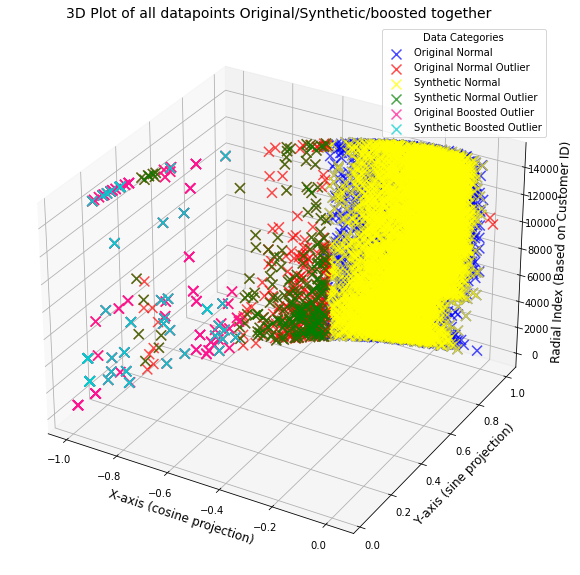} 
\caption{}
\label{Original_synthtic_boosted_postall}
\end{subfigure}\hfill
\caption{Figure showing the idea of angular distribution with blue region showing original data, yellow region being synthetic data, red region showing original data outlliers, green region showing synthetic data outleirs and finally cyan and pink showing boosted data from outliers of original or synthetic data. (a) Angular distribution and outlier regions for original data. (b) Angular distribution and outlier regions original and synthetic data separately. (c) Angular distribution and outlier regions original, synthetic and boosted data separately. (d) Angular distribution and outlier regions original, synthetic, and boosted data together.}
\label{qsmote_distribution_outliers}
\end{figure}

\subsection{Centroid Based Angular Outlier Boosting}\label{CAOL}

The concept of Angular Outliers was first introduced in Section 2. This study uses the Quantum-SMOTEV2 method to determine the angles between the data centroid and minority data points. Outliers are identified when they exceed the upper or lower limit set at $\pm 1.5$ times the interquartile range (IQR), known as Angular Outliers. This section evaluates the impact of enhancing Angular Outliers to improve classification efficiency across different SMOTE ratios, involving a two-step methodological approach. To further clarify this concept, we have included some figures that are briefly explained below.

\textbf{Figure \ref{Angular_Distribution}: Angular Distribution of Data} \
This histogram depicts the angular distances from the data centroid for minority data points. The thresholds for upper and lower outliers are shown by green and red dashed lines, respectively. Data points above these thresholds are designated as Angular Outliers. This visualization helps to identify the extent and distribution of outliers with regard to minority data points.

\textbf{Figure \ref{Original_outlier} (Original Data and Outliers)} \
This figure depicts the three-dimensional arrangement of the original minority data points. Red crosses denote outliers determined by the specified angular distance, highlighting their spatial segregation from the main cluster of blue points.

\textbf{Figure \ref{Original_synthtic_outlier} (Synthetic, Original, and Outliers)}\
This figure extends on the figure \ref{Original_outlier} by integrating synthetic data points, shown in yellow. It emphasizes the effectiveness of the data rotation technique used to generate synthetic data points, with synthetic outliers shown in green.

\textbf{Figure \ref{Original_synthtic_boosted_outlier} (Synthetic, Original, and Boosted Outliers)}\
 This figure shows the effect of the boosting technique on both original and synthetic datasets. The original data (red and blue dots) and synthetic data (green and yellow points) have been layered, with newly boosted data points shown in pink and light blue. These colors illustrate how boosted data points only appear in outlier regions above red(original) and green(synthetic).

\textbf{Figure \ref{Outlier_Distribution_2d}: Distribution of Angular Distance Outliers} \ This histogram shows the bins in the outlier region prior to boosting.

\textbf{Figure \ref{Original_synthtic_boosted_postall}: Post algorithm Data Distribution in 3D}
This graphic illustrates the actual distribution after the implementation of the boosting process.

\subsubsection{Algorithmic Implementation}

\textbf{Algorithm \ref{algo_outliers}}: This algorithm outlines the procedure of creating datasets from angular outliers based on the thresholds described. Two separate datasets are created: one for points beyond the upper threshold and another for those below the lower threshold. These datasets are further segmented by the 'bin' hyper-parameter, which governs the granularity of our analysis.

\textbf{Algorithm \ref{algo_boost}}: This algorithm outlines the process for enhancing underrepresented outlier bins. The enhancement procedure includes:
\begin{itemize}
    \item Counting the total entries in each outlier dataset.
    \item Establishing a threshold by dividing this count by the number of bins.
    \item Determining a half-threshold to identify bins with counts below half the average, which are then targeted for data augmentation.
    \item Utilizing broader rotation angles during boosting to avoid duplication of records and ensure a diverse data augmentation.
\end{itemize}

\subsection{Quantum-SMOTEV2 with Angular Outliers} \label{Approach}
In this paper, we present a variant of QuantumSMOTE with a refined algorithm that operates in three stages: First, we compute the data centroid. Second, synthetic data is generated for the minority class to achieve the desired minority proportion. Third, outliers are identified, and a certain percentage of them is amplified.

We introduce a modification in the second step. Previously, the algorithm calculated the angle between the centroid and a minority data point, followed by rotating the data point to generate synthetic data in a single operation. In this version, we split the process for better control. First, we calculate the angle between the centroid and all minority data points. Then, in the second step, we rotate each (or a chosen proportion) of the minority data points to generate synthetic data. This approach enables more precise management of synthetic data generation to meet specific target percentages.

After we create the synthetic data, our next steps include identifying and enhancing the outliers, as explained in more detail in Section \ref{CAOL}.

We've outlined the entire process in a pseudocode format in the next section. It consists of seven key phases: setting up the data for the swap test \ref{algo_swaptest_prep}, conducting the swap test itself \ref{algo_swaptest}, rotating synthetic data points \ref{algo_synthetic}, creating new synthetic points \ref{algo_syntheticgen}, spotting angular outliers \ref{algo_outliers}, and amplifying these outliers \ref{algo_boost}. Since we are building on an earlier version of this algorithm, some steps like preparing the data for the swap test, carrying out the swap test, and the methods we use for normalizing and rotating data remain unchanged.

\begin{algorithm}[H]
\caption{Preparation for Swap Test \cite{Mohanty24_Q-SMOTE}} 
\label{algo_swaptest_prep}
\begin{algorithmic}[1]
\Function{Prepswap test}{$data\_point1$, $data\_point2$}
    \State $norm\_data\_point1 \gets 0$
    \State $norm\_data\_point2 \gets 0$
    \State $Dist \gets 0$
    \For{$i \gets 0$ \textbf{to} $length(data\_point1)-1$}
        \State $norm\_data\_point1 \gets norm\_data\_point1 + data\_point1[i]^2$
        \State $norm\_data\_point2 \gets norm\_data\_point2 + data\_point2[i]^2$
        \State $Dist \gets Dist + (data\_point1[i] + data\_point2[i])^2$
    \EndFor
    \State $Dist \gets \sqrt{Dist}$
    \State $data\_point1\_norm \gets \sqrt{norm\_data\_point1}$
    \State $data\_point2\_norm \gets \sqrt{norm\_data\_point2}$
    \State $Z \gets \text{round}(data\_point1\_norm^2 + data\_point2\_norm^2)$
    \State $\phi \gets [\text{round}(data\_point1\_norm / \sqrt{Z}, 3), -\text{round}(data\_point2\_norm / \sqrt{Z}, 3)]$
    \State Initialize array $\psi$
    \For{$i \gets 0$ \textbf{to} $length(data\_point1)-1$}
        \State $\psi.\text{append}(\text{round}(data\_point1[i] / (data\_point1\_norm \times \sqrt{2}), 3))$
        \State $\psi.\text{append}(\text{round}(data\_point2[i] / (data\_point2\_norm \times \sqrt{2}), 3))$
    \EndFor
    \State \Return $\phi, \psi$
\EndFunction
\end{algorithmic}
\end{algorithm}

\begin{algorithm}[H]
\caption{Swap Test  \cite{Mohanty24_Q-SMOTE}}
\label{algo_swaptest}
\begin{algorithmic}[1]
\Function{swap testV1}{$\psi$, $\phi$}
    \State Initialize Quantum Register $q1$ with 1 qubit
    \State Initialize Quantum Register $q2$ with n+2 qubits
    \State Initialize Classical Register $c$ with 1 bit
    \State Create Quantum Circuit with $q1$, $q2$, and $c$

    \textbf{States initialization}
    \State Initialize $q2[0]$ with state $\phi$
    \State Initialize $q2[1:n+2]$ with state $\psi$

    \textbf{The swap test operator}
    \State Apply Pauli-X Gate to $q2[1]$

    \textbf{Swap Test}
    \State Apply Hadamard Gate to $q1[0]$
    \State Apply Controlled SWAP Gate on $q1[0]$, $q2[0]$, and $q2[1]$
    \State Apply Hadamard Gate to $q1[0]$
    \State Measure $q1$ into classical register $c$

    \textbf{Simulation and result collection}
    \State Set up quantum simulator
    \State Execute the quantum circuit on the simulator
    \State Collect the result into a variable $result$
    \State Extract measurement counts from $result$

    \textbf{Calculate the Swap Test probability}
    \State $p0 \gets \frac{\text{counts.get('0', 0)}}{total\_shots}$
    \State $p1 \gets \frac{\text{counts.get('1', 0)}}{total\_shots}$
    \State $swap\_test\_probability \gets 1 - 2 \times p0 + p1$
    \State Print $swap\_test\_probability$

    \textbf{Calculate the angular distance}
    \State $angular\_distance \gets 2 \times \text{arccos}(\sqrt{swap\_test\_probability})$
    \State Print $angular\_distance$

    \State \Return $swap\_test\_probability$, $angular\_distance$
\EndFunction
\end{algorithmic}
\end{algorithm}

\begin{algorithm}
\caption{Normalize Array \cite{Mohanty24_Q-SMOTE}}[H]
\label{algo_Normalize_Array}
\begin{algorithmic}[1]
\Function{NormalizeArray}{$arr$}
    \textbf{Calculate the sum of squares of the elements in the array}
    \State $sum\_of\_squares \gets \Call{SumOfSquares}{arr}$

    \textbf{Check if the sum of squares is already very close to 1}
    \If{\Call{IsClose}{$sum\_of\_squares$, $1.0$, $\text{rtol}=1e-6$}}
        \State \Return $arr$
    \EndIf

    \textbf{Calculate the scaling factor to make the sum of squares equal to 1}
    \State $scaling\_factor \gets 1.0 / \sqrt{sum\_of\_squares}$

    \textbf{Normalize the array by multiplying each element by the scaling factor}
    \State $normalized\_arr \gets arr \times scaling\_factor$

    \State \Return $normalized\_arr$
\EndFunction
\end{algorithmic}
\end{algorithm}

\begin{algorithm}[H]
\caption{Create Synthetic Data}
\label{algo_syntheticgen}
\begin{algorithmic}[1]
\State \textbf{Input:} $n$ (number of qubits), $angle\_increment$, $angular\_distance$, $data\_point1$
\State \textbf{Output:} $new\_data\_point$, $angle$

\Function{CreateSynData}{$n$, $angle\_increment$, $angular\_distance$, $data\_point1$}
    \State Normalize $data\_point1$
    \State Initialize quantum circuit with $n$ qubits
    \State Apply $data\_point1$ to initialize the quantum circuit

    \If {$angular\_distance > \frac{\pi}{2}$}
        \State $angle \gets \frac{\left| \frac{\pi}{2} - angular\_distance \right|}{10}$
    \ElsIf {$angular\_distance < 0$}
        \State $angle \gets \frac{\left| \frac{\pi}{2} - angular\_distance \right| \cdot \text{RandomUniform}(0.5, 1)}{10}$
    \Else
        \State $angle \gets \frac{\text{RandomUniform}(0, angular\_distance)}{10}$
    \EndIf

    \State $angle \gets angle + angle\_increment$
    \State \textbf{Print} "rotation angle", $angle$

    \For {$l \gets 0$ \textbf{to} $n-1$}
        \State Apply RX gate $R_x(angle)$ to qubit $l$
    \EndFor

    \State Simulate the quantum circuit using a statevector simulator
    \State Extract $statevector$ from the simulation results

    \State $new\_data\_point \gets \text{Real}(statevector)$
    \State \Return $new\_data\_point$, $angle$
\EndFunction
\end{algorithmic}
\end{algorithm}

\begin{algorithm}[H]
\caption{Synthetic Data Creation Process Quantum-SMOTEV2}
\label{algo_synthetic}
\begin{algorithmic}[1]
\State \textbf{Input:} $n$ (number of qubits), $target\_synthetic\_percent$, $minority\_set$, $centroid\_df\_row$
\State \textbf{Output:} Synthetic data for minority class

\Function{CalculateAngle}{$minority\_dp$, $centroid\_dp$}
    \State Normalize $minority\_dp$ and $centroid\_dp$
    \State Apply \textit{swap test} to calculate $swap\_test\_probability$, $angular\_distance$
    \State \Return $swap\_test\_probability$, $angular\_distance$
\EndFunction

\Function{CreateSyntheticData}{$n$, $minority\_set$, $centroid\_dp$, $target\_synthetic\_percent$, $angular\_distance$}
    \State Compute $minority\_count \gets \text{count of minority class in the dataset}$
    \State Compute $total\_count \gets \text{total count of records in the dataset}$
    \State Compute $minority\_percent \gets \left( \frac{minority\_count}{total\_count} \right) \times 100$
    
    \State Compute $target\_minority\_count \gets total\_count \times \frac{target\_synthetic\_percent}{100}$
    \State Compute $target\_synthetic\_count \gets \frac{target\_minority\_count - minority\_count}{1 - \frac{target\_synthetic\_percent}{100}}$
    \State Compute $synthetic\_loop\_itr \gets \frac{target\_synthetic\_count}{minority\_count}$
    \State Compute $rem\_synthetic\_loop\_itr \gets \text{mod}(target\_synthetic\_count, minority\_count)$

    \State Initialize $syn\_dataframe$ with required columns for synthetic data
    \For {$syn\_loop \gets 1$ to $synthetic\_loop\_itr$}
        \If {$syn\_loop == synthetic\_loop\_itr-1$}
            \State Select $minority\_temp \gets \text{randomly sample remaining minority data points}$
        \Else
            \State Set $minority\_temp \gets minority\_set$
        \EndIf

        \For {each minority data point $dp$ in $minority\_temp$}
            \State Retrieve $angular\_distance$ for each data point using \textit{CalculateAngle} function
            \State Calculate $n \gets \log_2(len(dp))$
            \State Set $loop\_ctr \gets \frac{len(dp)}{n}$ and round to the nearest integer
            \State Set $angle\_increment \gets syn\_loop \times 0.0174533$

            \State Generate synthetic data $syn\_data$ and $rotation\_angle$ using \textit{CreateSynData} function
            \State Append synthetic data $syn\_df\_temp$ to $syn\_dataframe$
        \EndFor
    \EndFor

    \State \Return $syn\_dataframe$ containing all synthetic data points
\EndFunction

\end{algorithmic}
\end{algorithm}

\begin{algorithm} [H]
\caption{Generate Outlier Datasets after Appending Synthetic Data}
\label{algo_outliers}
\begin{algorithmic}[1]
\State \textbf{Input:} $Minority\_Df\_Orig$, $syn\_dataframe$, $num\_bins$
\State \textbf{Output:} $outlier\_low\_bins\_df$, $outlier\_high\_bins\_df$

\Function{GenerateOutliers}{$Minority\_Df\_Orig$, $syn\_dataframe$, $num\_bins$}
    \State Concatenate $Minority\_Df\_Orig$ and $syn\_dataframe$ into $Minority\_synthetic\_df$
    \State Calculate $Q1 \gets \text{25th percentile of } angular\_distance$ in $Minority\_synthetic\_df$
    \State Calculate $Q3 \gets \text{75th percentile of } angular\_distance$ in $Minority\_synthetic\_df$
    \State Calculate Interquartile Range (IQR): $IQR \gets Q3 - Q1$
    
    \State Set outlier thresholds:
    \State $lower\_bound \gets Q1 - 1.5 \times IQR$
    \State $upper\_bound \gets Q3 + 1.5 \times IQR$
    
    \State Identify outliers based on $angular\_distance$:
    \State $outliers\_low \gets \text{Subset of data with } angular\_distance < lower\_bound$
    \State $outliers\_high \gets \text{Subset of data with } angular\_distance > upper\_bound$
    
    \State Group outliers into bins:
    \State $outliers\_low\_bins \gets \Call{HistogramBinEdges}{outliers\_low['angular\_distance'], num\_bins}$
    \State $outlier\_low\_counts, \_ \gets \Call{Histogram}{outliers\_low['angular\_distance'], outliers\_low\_bins}$
    
    \State $outliers\_high\_bins \gets \Call{HistogramBinEdges}{outliers\_high['angular\_distance'], num\_bins}$
    \State $outlier\_high\_counts, \_ \gets \Call{Histogram}{outliers\_high['angular\_distance'], outliers\_high\_bins}$

    \State Create a DataFrame of low outlier bins and counts:
    \State $outlier\_low\_bins\_df \gets \Call{DataFrame}{\{'Bin\_Start': outliers\_low\_bins[:-1], 'Bin\_End': outliers\_low\_bins[1:], 'Count': outlier\_low\_counts\}}$
    
    \State Create a DataFrame of high outlier bins and counts:
    \State $outlier\_high\_bins\_df \gets \Call{DataFrame}{\{'Bin\_Start': outliers\_high\_bins[:-1], 'Bin\_End': outliers\_high\_bins[1:], 'Count': outlier\_high\_counts\}}$
    
    \State \Return $outlier\_low\_bins\_df$, $outlier\_high\_bins\_df$
\EndFunction

\end{algorithmic}
\end{algorithm}

\begin{algorithm} [H]
\caption{Boosting Outlier Dataset using Quantum-SMOTEV2}
\label{algo_boost}
\begin{algorithmic}[1]
\State \textbf{Input:} $outlier\_df$, $smote\_ds$, $target\_col$, $target\_col\_val$, $num\_bins$
\State \textbf{Output:} $boost\_syn\_dataframe$ with boosted synthetic data

\Function{QuantumSMOTEBoost}{$outlier\_df$, $smote\_ds$, $target\_col$, $target\_col\_val$, $num\_bins$}
    \State Calculate $total\_outlier\_recs \gets \Call{Sum}{outlier\_df['Count']}$
    \State Calculate $threshold \gets \Call{Round}{total\_outlier\_recs / num\_bins}$
    \State Calculate $half\_threshold \gets \Call{Round}{threshold / 2}$
    
    \State Initialize $boost\_syn\_dataframe$ with columns from $smote\_ds$
    \State Add additional columns: $Boosted$, $Rotation\_angle$
    
    \For {$i \gets 0$ \textbf{to} $len(outlier\_df) - 1$}
        \If {$outlier\_df['Count'][i] < half\_threshold$}
            \State Set $local\_bin\_count \gets outlier\_df['Count'][i]$
            
            \State Retrieve $minority\_temp \gets \text{subset of smote\_ds where angular\_distance is within the bin range}$
            
            \State Calculate $synthetic\_loop\_itr \gets \Call{Floor}{threshold / local\_bin\_count}$
            
            \For {each $row$ in $minority\_temp$}
                \State $minority\_dp\_temp \gets row[minority\_temp.columns[:-8]]$
                \State $n \gets \log_2(len(minority\_dp\_temp))$
                \State Adjust $n$ based on whether the length is divisible by $n$
                
                \For {$j \gets 0$ \textbf{to} $synthetic\_loop\_itr - 1$}
                    \State Set $angle\_increment \gets (synthetic\_loop\_itr \times 0.0174533) \times 1.5 + j$
                    \State Set $angular\_distance \gets row['angular\_distance']$
                    
                    \State Call \texttt{CreateSynData} with $n$, $angle\_increment$, $angular\_distance$, and the normalized $minority\_dp\_temp$
                    \State Assign the output to $boost\_syn\_data$ and $rot\_angle$
                    
                    \State Create $boost\_syn\_df\_temp$ from $boost\_syn\_data$ and set additional metadata fields
                    \State Set $boost\_syn\_df\_temp['Boosted'] \gets 'Yes'$
                    \State Set $boost\_syn\_df\_temp['Rotation\_angle'] \gets rot\_angle$
                    
                    \State Append $boost\_syn\_df\_temp$ to $boost\_syn\_dataframe$
                \EndFor
            \EndFor
        \EndIf
    \EndFor
    
    \State \Return $boost\_syn\_dataframe$
\EndFunction

\end{algorithmic}
\end{algorithm}

\section{Case Study and Results} \label{Case Study and Results}
We evaluate the Quantum-SMOTEV2 method by analyzing the publicly accessible telecom churn dataset \cite{telco_churn_dataset_kaggle}. This dataset is extensively used for experimenting with and evaluating different customer retention models, proving valuable for comparing traditional models with those enhanced by the Quantum-SMOTEV2 algorithm's synthetic data induction. Subsequent sections will discuss data behavior, data preparation for modeling, and the application of Quantum-SMOTEV2 to the data.

\subsection{Telecom Churn Prediction Using Q-SMOTE-AOL} \label{Telecom Churn Prediction Using Q-SMOTE-AOL}

The cell-to-cell telecommunications churn dataset is specifically designed for predicting customer behaviour and assist in the formulation of customer retention strategies. Each row in this set signifies a distinct consumer, whereas each column denotes various properties of these customers. It contains 51,047 entries and 58 characteristics about consumer behavior and subscriptions. Below are its main features:

\textbf{Customer:} Customers have unique CustomerID and demographic information like age and if they have children.
\textbf{Service Usage:} Monthly income, minutes utilised, total recurring costs, overage minutes, and call kinds (dropped, blocked, unanswered) are reported.
\textbf{Account Changes:} The dataset captures client account changes, including phone types, equipment days, and service area.
\textbf{Engagement Metrics:} Customer care, three-way, and roaming calls reveal customer engagement.
\textbf{Retention metrics:} Telephone calls, offers accepted, and subscriber referrals are important for churn research.
\textbf{Financial metrics:} Monthly revenue and credit rating changes may indicate client satisfaction and turnover.

\subsubsection{Preparing Data }
The Telco churn dataset is suitable for a standard data preparation procedure, which generally encompasses the following steps.

\textbf{Exploratory Data Analysis}: EDA was performed to understand the distribution and relationship of variables. We applied various univariate and bivariate statistics as well as correlation analysis to effectively judge relationships between variables and eliminate multicolinear features.

\textbf{Removing Irrelevant Data}: We have carefully selected relevant columns that are important for churn prediction and dropped several colummns such as MonthlyMinutes, PercChangeRevenues, ReceivedCalls, and CurrentEquipmentDays.

\textbf{Data cleaning}: We have applied standard data cleaning techniques that included missing value treatment, data type conversion and dropping irrelevant columns to prepare features for Modelling.

\textbf{Binning and label encoding}
To deal with numerical columns that are of different distribution, we have binned several columns into discrete interval bins, which are derived from the actual data ranges. This offers advantages like simplification, handling non-linear relationships, improving robustness, reducing overfitting etc. These bins are further label encoded into numerical values to simplify the overall process. Finally we selected the folloing columns

\begin{itemize}
    \item \textbf{ID Column} 
    'CustomerID', 
    \item \textbf{Categorical Columns} 
'HandsetModels', 'ChildrenInHH', 'HandsetRefurbished', 'HandsetWebCapable', 
 'TruckOwner', 'RVOwner', 'Homeownership',  'BuysViaMailOrder',  'RespondsToMailOffers',  'OptOutMailings', 'NonUSTravel',
 'OwnsComputer',  'HasCreditCard', 'RetentionCalls', 'RetentionOffersAccepted', 'NewCellphoneUser',  'NotNewCellphoneUser',  'IncomeGroup', 'OwnsMotorcycle',  'MadeCallToRetentionTeam',  'CreditRating', 'PrizmCode', 'Occupation',   'MaritalStatus',
 \item \textbf{Binned Numerical Columns} 
 'MonthlyRevenue\_Bin', 'TotalRecurringCharge\_Bin','DirectorAssistedCalls\_Bin', 'OverageMinutes\_Bin', 'RoamingCalls\_Bin', 'PercChangeMinutes\_Bin', 'DroppedCalls\_Bin', 'UnansweredCalls\_Bin', 'CustomerCareCalls\_Bin', 
'ThreewayCalls\_Bin',  'OutboundCalls\_Bin', 'InboundCalls\_Bin', 'PeakCallsInOut\_Bin', 'OffPeakCallsInOut\_Bin', 
'DroppedBlockedCalls\_Bin',  'CallForwardingCalls\_Bin', 'CallWaitingCalls\_Bin', 'MonthsInService\_Bin', 'UniqueSubs\_Bin',  'ActiveSubs\_Bin',  'Handsets\_Bin',
 'AgeHH1\_Bin', 'HandsetPrice\_Bin', 'AdjustmentsToCreditRating\_Bin', 'ReferralsMadeBySubscriber\_Bin', 
 \item \textbf{Target Column}   'Churn'
\end{itemize}

\subsubsection{Applying SMOTE and Outlier Boost on Prepared Data}
For the data preparation of the cell-2-cell dataset and  we proceeded to apply our proposed Quantum-SMOTEV2 (\ref{algo_syntheticgen}) to the entire dataset. The objective was to steadily enhance the representation of the minority population to a certain proportion of the entire dataset and thereafter implement the amplification of angular outliers.The procedure is carried out by gradually increasing the minority percentage from 28.5\% to all the way up to 50\% with each step involving an outlier boost. The procedure used two primary approaches previously mentioned, namely the Quantum-SMOTEV2 (Algo. \ref{algo_syntheticgen}) and Outlier boost (Algo. \ref{algo_boost}). 

\begin{figure}[H]
    \centering
    \includegraphics[width=0.6\linewidth]{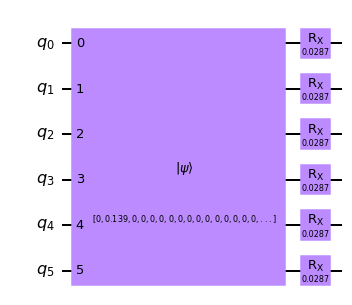}
    \caption{Data point rotation circuit.}
\label{qsmote_rotationcircuit}
\end{figure}

\textbf{Swap Test and Rotation}: In our previous paper \cite{Mohanty24_Q-SMOTE} we have explained the use of compact swap test \cite{qiskit_medium, Mart_Dissimilarity_2023} followed by rotation for generating synthetic samples. However, in this paper, we have an additional step to boost a portion of outliers using the same principles but with a wider rotation angle to avoid duplicates. The difference in this paper is we are using a single data centroid instead of multiple cluster centroid. We have used similar circuits for the compact swap test that is obtained is rendered in Fig. \ref{Compact Swap test circuit.}. Similarly, the quantum rotation follows identical circuits \ref{qsmote_rotationcircuit}.

The primary advantages of using a compact swap test circuit in reducing the number of qubits still remain intact, as well as the advantages of quantum rotation.

\subsubsection{Applying Classification Models Accessing Impact}
To provide an overall effectiveness test of Quantum-SMOTEV2 and the boosting of outliers in handling class imbalances, we applied three classification models: RF, KNN, and NN on the dataset of cell to cell. These models have been chosen to see the effect of using Quantum-SMOTEV2 and boosting outliers in order to improve their performance, especially when applied to a highly imbalanced class distribution problem. The RF algorithm is widely known for its ability to handle highly biased data with high efficiency. This model employs ensemble learning by generating several decision trees, after which it pools predictions to avoid overfitting. The algorithm inherently addresses class imbalances using methods like bootstrap sampling and adjusts its class weights parameter to enhance sensitivity to the minority class. In many cases, RF does not require external interventions such as SMOTE \cite{breiman_random_2001}.

KNN, has one of the most simple and intuitive approaches toward performing classification. It is particularly well-suited to data nonlinearly-separable. After Cover and Hart \cite{Cover_Hart_KNN_1967}, the two authors of the KNN algorithm, the principle is to get the k nearest data points in the feature space. A new instance will be classified according to the majority vote among those classes. This non-parametric technique adapts to the underlying distribution in a flexible way but is sensitive when the dataset is unbalanced. The elementary techniques that help increase the robustness of this class include varying k and distance-weighting described by Hechenbichler and Schliep \cite{Hechenbichler_Schliep_KNN_2004}. However, most of these schemes tend to be sensitive and require proper normalization of input features and choice of the distance metric to avoid skewness of the class distribution. 

NNs are quite flexible and powerful models that can capture higher-order data relationships due to an increase in the number of layers and non-linear activation functions. According to \cite{LeCun_Bengio_Hinton_2015}, these networks tune their inner parameters (weights) in a procedure called backpropagation that updates weights such that prediction error is minimized. NNs can be suitably adapted to handle challenges in imbalanced datasets either using cost-sensitive learning or by modifying the objective function to focus on the minority classes \cite{Buda_Maki_NN_2018}. These adaptations may make the model sensitive to under-represented data points, and therefore, highly useful for a wide range of applications out there, from recognizing images to predicting consumer behaviors. In this research, we have used a Deep NN to initially evaluate model performance with synthetic data and then with induction of synthetic samples in a gradually increasing manner from 30\% to 50\% along with Outlier boosting. The NN used in this paper consists of 4 dense layers (128,64,64,32) with Relu as an activation function, with the final layer being sigmoid, and is trained on 50 epochs with batch size 32 and features such as early stopping.

For the sake of readers' convenience, the initial models are biased with lots of false negatives, which gradually improves upon the addition of synthetic data. We have used Confusion Matrix, Accuracy, Precision, Recall, F1-Score, and the Area Under the Receiver Operating Characteristic Curve (AUC-ROC) to assess models. Presented here \ref{Model Performance} are the model assessment charts for the RF Model, followed by KNN Classification and NN.

\begin{figure}[H]
    \centering
    \includegraphics[width=0.8\linewidth]{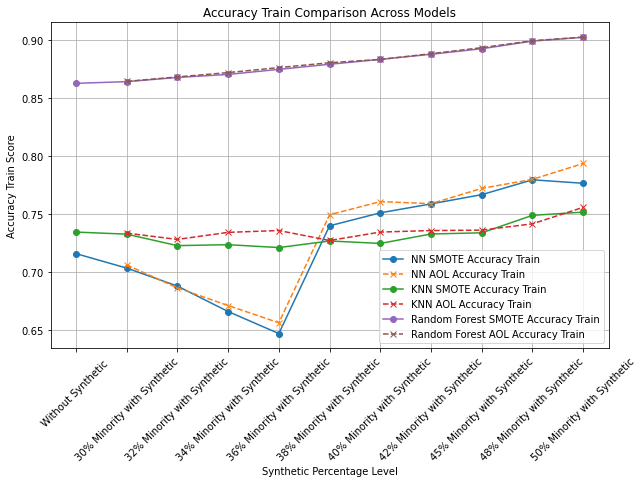}
    \caption{Figure Showing train accuracy across RF, NN, KNN with/without outlier boosting for varying levels of Quantum-SMOTEV2 }
    \label{Train_Accuracy_Trend}
\end{figure}

\begin{figure}[H]
    \centering
    \includegraphics[width=0.8\linewidth]{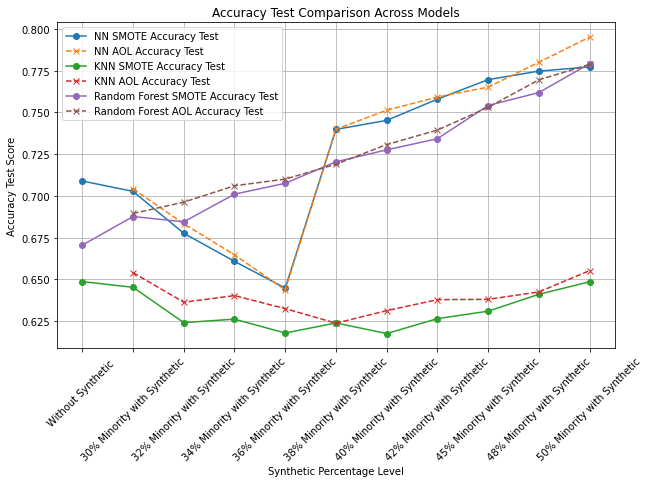}
    \caption{Figure Showing test accuracy across RF, NN, KNN with/without outlier boosting for varying levels of Quantum-SMOTEV2}
    \label{Test_Accuracy_Trend}
\end{figure}

\begin{figure}[H]
    \centering
    \includegraphics[width=0.8\linewidth]{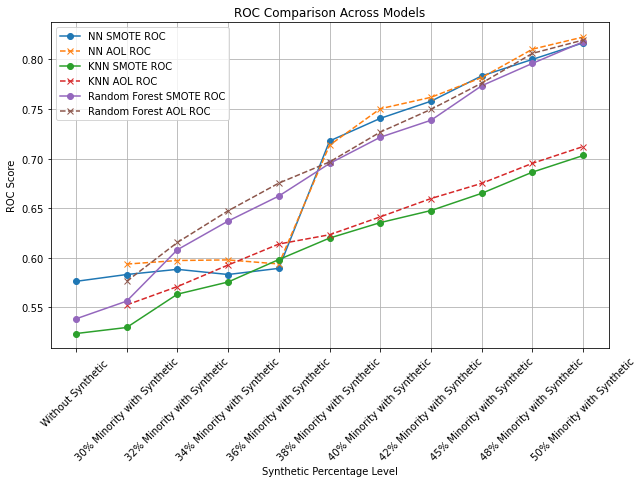}
    \caption{Figure Showing ROC-AUC across RF, NN, KNN with/without outlier boosting for varying levels of Quantum-SMOTEV2}
    \label{ROC_AUC_Trend}
\end{figure}

\begin{figure}[H]
    \centering
    \includegraphics[width=0.8\linewidth]{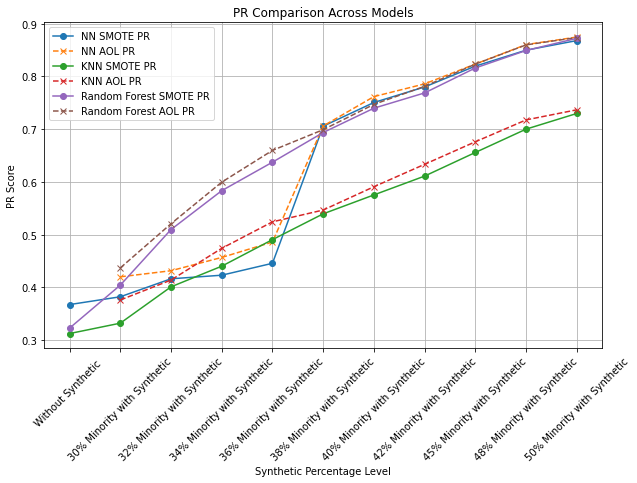}
    \caption{Figure Showing Precision-Recall across RF, NN, KNN with/without outlier boosting for varying levels of Quantum-SMOTEV2}
    \label{PR_AUC_Trend}
\end{figure}

\begin{figure}[H]
    \centering
    \includegraphics[width=0.8\linewidth]{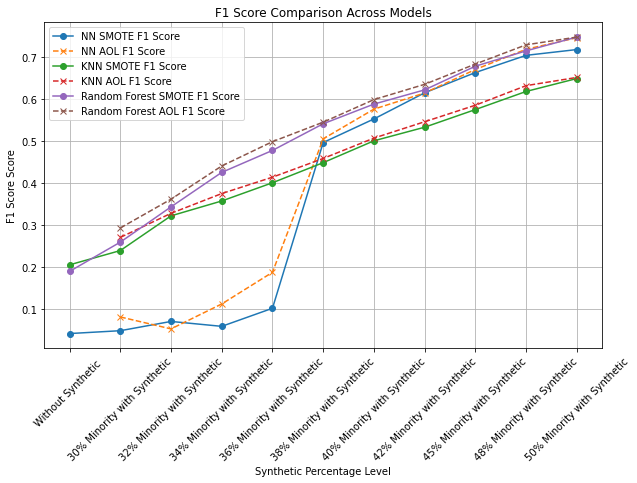}
    \caption{Figure Showing F1 Score across RF, NN, KNN with/without outlier boosting for varying levels of Quantum-SMOTEV2}
    \label{F1_Score_Trend}
\end{figure}

As we applied SMOTE to our three chosen models, we observed different behaviors of the models post-application of QuantumSMOTE. We charted the model evaluation parameters for varying levels of smote in figures \ref{Train_Accuracy_Trend}, \ref{Test_Accuracy_Trend}, \ref{ROC_AUC_Trend}, \ref{PR_AUC_Trend} and \ref{F1_Score_Trend}. We summarize our observations in the next section \ref{Model_Evaluation}. 

\section{Model evaluation} \label{Model_Evaluation}

\subsubsection{Accuracy}
In terms of accuracy, all three models demonstrated improvements with the introduction of SMOTE and Outlier Boosting. Initially, without any synthetic data, RF outperformed both NN and KNN, achieving a test accuracy of 0.671. However, as SMOTE percentages were gradually increased, RF maintained its lead, with accuracy improving from 0.688 at 30\% SMOTE to 0.779 at 50\% SMOTE.

NN showed significant gains in accuracy as SMOTE percentages increased, starting from 0.703 at 30\% SMOTE and reaching 0.777 at 50\%. Outlier Boosting further enhanced its performance, allowing it to reach its peak accuracy at 50\% SMOTE.

KNN, on the other hand, demonstrated more stable, moderate improvements. Its accuracy showed minimal fluctuations, from 0.645 at 30\% SMOTE to 0.649 at 50\% SMOTE. This indicates that KNN, while improving with synthetic data, is less sensitive to higher SMOTE percentages compared to NN and RF.

\textbf{Interpretation}: RF exhibits superior performance in terms of accuracy across all settings, particularly benefiting from higher SMOTE percentages and Outlier Boosting. NN demonstrates a notable rise in accuracy as SMOTE is increased, while KNN achieves only marginal improvements.

\subsubsection{ROC AUC}
ROC AUC, which measures the model’s ability to differentiate between classes, showed significant improvement across all models when synthetic data was introduced. RF exhibited the highest baseline ROC AUC of 0.539 without SMOTE, which steadily improved as SMOTE was increased. By 50\% SMOTE, RF reached its peak ROC AUC of 0.818, highlighting its capacity for handling imbalanced data with the help of synthetic data and Outlier Boosting.

NN started with a lower baseline ROC AUC of 0.576. However, the model’s ability to distinguish between classes improved significantly with SMOTE, reaching 0.817 at 50\% SMOTE, almost equal to RF.

KNN, starting at a lower baseline of 0.524, showed moderate improvements, with its ROC AUC increasing to 0.703 at 50\% SMOTE. This indicates that while KNN does benefit from the addition of synthetic data, its gains in ROC AUC are less pronounced compared to NN and RF.

\textbf{Interpretation}: RF and NN both show substantial improvements in ROC AUC with increasing SMOTE percentages and Outlier Boosting, indicating a better capacity for class separation. KNN, while improving, lags behind in terms of ROC AUC, suggesting that it is less effective at class differentiation in highly imbalanced datasets.

\subsubsection{Precision-Recall and  F1 Score}
F1 Score and PR AUC, which are critical metrics for imbalanced datasets, improved dramatically for all models with the introduction of SMOTE and Outlier Boosting. Initially, NN showed a very low F1 score (0.041) and PR AUC (0.368), indicating its poor handling of imbalanced data without synthetic data. However, with increasing SMOTE percentages, the model’s F1 score rose significantly, reaching 0.719 at 50\% SMOTE, alongside a PR AUC of 0.868. This showcases NN’s enhanced ability to manage imbalanced data through the addition of synthetic minority class examples.

KNN demonstrated a better baseline F1 score (0.205) and PR AUC (0.313) than NN, but it also benefited from SMOTE and Outlier Boosting. By 50\% SMOTE, KNN achieved an F1 score of 0.650 and PR AUC of 0.730, reflecting moderate gains in both metrics.

RF, which started with an F1 score of 0.190 and PR AUC of 0.323, demonstrated the most significant improvements as SMOTE increased. With Outlier Boosting and 50\% SMOTE, RF achieved the highest F1 score (0.748) and PR AUC (0.872), making it the best performer across all models in terms of precision-recall metrics.

\textbf{Interpretation}:RF exhibits the most robust performance in handling imbalanced data, as indicated by the highest F1 score and PR AUC with increasing SMOTE and Outlier Boosting. NN shows substantial improvements as well, making it a strong alternative, while KNN, though improved, lags behind the other models in precision-recall metrics.

The entire model evaluation parameters are tabulated in the table \ref{Model_Evaluation_Stats}

\begin{table}[ht]
\centering
\refstepcounter{table}
\resizebox{\linewidth}{!}{%
\begin{tabular}{|l|c|c|c|c|c|c|c|c|c|c|} 
\hline
\multicolumn{11}{|c|}{{\cellcolor[rgb]{0.855,0.949,0.816}}\textbf{RF}}                                                                                                                                                                                                                                            \\ 
\hline
\multicolumn{1}{|c|}{\textbf{Scores}} & \multicolumn{2}{c|}{\textbf{Accuracy Score}}               & \multirow{2}{*}{\textbf{F1 Score}}        & \multicolumn{2}{c|}{\textbf{AUC Score}}                                               & \multicolumn{2}{c|}{\textbf{Accuracy Score AOL}}                  & \multirow{2}{*}{\textbf{F1 Score AOL}}    & \multicolumn{2}{c|}{\textbf{AUC Score AOL}}                                            \\ 
\textbf{Data Set Type} & \textbf{Train} & \textbf{Test} & & \textbf{PR} & \textbf{ROC} & \textbf{Train} & \textbf{Test} & & \textbf{PR} & \textbf{ROC} \\ 
\hline
Without Synthetic                     & 0.863          & {\cellcolor[rgb]{0.988,0.988,1}}0.671     & {\cellcolor[rgb]{0.753,0.902,0.961}}0.190 & {\cellcolor[rgb]{0.949,0.808,0.937}}0.323 & {\cellcolor[rgb]{0.949,0.949,0.949}}0.539 & \multicolumn{1}{l|}{} & \multicolumn{1}{l|}{}                     &                                           &                                           &                                            \\ 
\hline
30\% Minority with Synthetic           & 0.864          & {\cellcolor[rgb]{0.894,0.953,0.922}}0.688 & {\cellcolor[rgb]{0.694,0.878,0.953}}0.259 & {\cellcolor[rgb]{0.937,0.753,0.918}}0.404 & {\cellcolor[rgb]{0.949,0.949,0.89}}0.556  & 0.864                 & {\cellcolor[rgb]{0.988,0.988,1}}0.689     & {\cellcolor[rgb]{0.753,0.902,0.961}}0.293 & {\cellcolor[rgb]{0.949,0.808,0.937}}0.437 & {\cellcolor[rgb]{0.949,0.949,0.949}}0.576  \\ 
\hline
32\% Minority with Synthetic           & 0.868          & {\cellcolor[rgb]{0.914,0.961,0.937}}0.684 & {\cellcolor[rgb]{0.62,0.847,0.941}}0.343  & {\cellcolor[rgb]{0.918,0.682,0.894}}0.510 & {\cellcolor[rgb]{0.961,0.961,0.714}}0.608 & 0.868                 & {\cellcolor[rgb]{0.945,0.973,0.961}}0.696 & {\cellcolor[rgb]{0.682,0.875,0.949}}0.362 & {\cellcolor[rgb]{0.933,0.737,0.914}}0.521 & {\cellcolor[rgb]{0.957,0.957,0.8}}0.616    \\ 
\hline
34\% Minority with Synthetic           & 0.870          & {\cellcolor[rgb]{0.824,0.922,0.855}}0.701 & {\cellcolor[rgb]{0.549,0.82,0.929}}0.426  & {\cellcolor[rgb]{0.902,0.627,0.875}}0.584 & {\cellcolor[rgb]{0.965,0.965,0.616}}0.637 & 0.872                 & {\cellcolor[rgb]{0.878,0.945,0.906}}0.706 & {\cellcolor[rgb]{0.596,0.839,0.937}}0.441 & {\cellcolor[rgb]{0.914,0.667,0.89}}0.600  & {\cellcolor[rgb]{0.961,0.961,0.675}}0.647  \\ 
\hline
36\% Minority with Synthetic           & 0.875          & {\cellcolor[rgb]{0.784,0.906,0.827}}0.708 & {\cellcolor[rgb]{0.502,0.8,0.922}}0.478   & {\cellcolor[rgb]{0.894,0.592,0.863}}0.638 & {\cellcolor[rgb]{0.969,0.969,0.529}}0.662 & 0.876                 & {\cellcolor[rgb]{0.851,0.933,0.882}}0.710 & {\cellcolor[rgb]{0.533,0.812,0.925}}0.499 & {\cellcolor[rgb]{0.898,0.616,0.871}}0.660 & {\cellcolor[rgb]{0.969,0.969,0.565}}0.675  \\ 
\hline
38\% Minority with Synthetic           & 0.879          & {\cellcolor[rgb]{0.714,0.878,0.765}}0.720 & {\cellcolor[rgb]{0.447,0.776,0.914}}0.542 & {\cellcolor[rgb]{0.882,0.553,0.851}}0.694 & {\cellcolor[rgb]{0.976,0.976,0.42}}0.695  & 0.881                 & {\cellcolor[rgb]{0.792,0.91,0.831}}0.719  & {\cellcolor[rgb]{0.486,0.792,0.918}}0.545 & {\cellcolor[rgb]{0.89,0.58,0.859}}0.699   & {\cellcolor[rgb]{0.973,0.973,0.482}}0.697  \\ 
\hline
40\% Minority with Synthetic           & 0.883          & {\cellcolor[rgb]{0.675,0.863,0.729}}0.728 & {\cellcolor[rgb]{0.408,0.761,0.906}}0.589 & {\cellcolor[rgb]{0.875,0.522,0.839}}0.740 & {\cellcolor[rgb]{0.98,0.98,0.329}}0.721   & 0.883                 & {\cellcolor[rgb]{0.714,0.878,0.761}}0.731 & {\cellcolor[rgb]{0.427,0.769,0.91}}0.599  & {\cellcolor[rgb]{0.878,0.541,0.843}}0.747 & {\cellcolor[rgb]{0.98,0.98,0.365}}0.727    \\ 
\hline
42\% Minority with Synthetic           & 0.888          & {\cellcolor[rgb]{0.639,0.847,0.698}}0.734 & {\cellcolor[rgb]{0.376,0.749,0.902}}0.622 & {\cellcolor[rgb]{0.867,0.502,0.831}}0.769 & {\cellcolor[rgb]{0.984,0.984,0.271}}0.739 & 0.888                 & {\cellcolor[rgb]{0.655,0.855,0.714}}0.739 & {\cellcolor[rgb]{0.388,0.753,0.902}}0.636 & {\cellcolor[rgb]{0.871,0.51,0.835}}0.781  & {\cellcolor[rgb]{0.984,0.984,0.275}}0.750  \\ 
\hline
45\% Minority with Synthetic           & 0.893          & {\cellcolor[rgb]{0.529,0.804,0.604}}0.754 & {\cellcolor[rgb]{0.329,0.729,0.894}}0.679 & {\cellcolor[rgb]{0.859,0.467,0.82}}0.816  & {\cellcolor[rgb]{0.988,0.988,0.153}}0.773 & 0.894                 & {\cellcolor[rgb]{0.565,0.816,0.631}}0.753 & {\cellcolor[rgb]{0.337,0.733,0.894}}0.684 & {\cellcolor[rgb]{0.859,0.475,0.82}}0.823  & {\cellcolor[rgb]{0.988,0.988,0.173}}0.776  \\ 
\hline
48\% Minority with Synthetic           & 0.899          & {\cellcolor[rgb]{0.486,0.784,0.565}}0.762 & {\cellcolor[rgb]{0.298,0.718,0.89}}0.716  & {\cellcolor[rgb]{0.855,0.447,0.812}}0.849 & {\cellcolor[rgb]{0.992,0.992,0.075}}0.796 & 0.900                 & {\cellcolor[rgb]{0.451,0.773,0.537}}0.769 & {\cellcolor[rgb]{0.29,0.714,0.886}}0.730  & {\cellcolor[rgb]{0.851,0.443,0.812}}0.860 & {\cellcolor[rgb]{0.996,0.996,0.055}}0.806  \\ 
\hline
50\% Minority with Synthetic           & 0.903          & {\cellcolor[rgb]{0.388,0.745,0.482}}0.779 & {\cellcolor[rgb]{0.267,0.702,0.882}}0.748 & {\cellcolor[rgb]{0.847,0.427,0.804}}0.872 & {\cellcolor{yellow}}0.818                 & 0.903                 & {\cellcolor[rgb]{0.388,0.745,0.482}}0.779 & {\cellcolor[rgb]{0.267,0.702,0.882}}0.748 & {\cellcolor[rgb]{0.847,0.427,0.804}}0.874 & {\cellcolor{yellow}}0.819                  \\ 
\hline
\multicolumn{11}{|c|}{{\cellcolor[rgb]{0.302,0.576,0.851}}\textbf{KNN Classifier}}                                                                                                                                                                                                                                            \\ 
\hline
\multicolumn{1}{|c|}{\textbf{Scores}} & \multicolumn{2}{c|}{\textbf{Accuracy Score}}               & \multirow{2}{*}{\textbf{F1 Score}}        & \multicolumn{2}{c|}{\textbf{AUC Score}}                                               & \multicolumn{2}{c|}{\textbf{Accuracy Score AOL}}                  & \multirow{2}{*}{\textbf{F1 Score AOL}}    & \multicolumn{2}{c|}{\textbf{AUC Score AOL}}                                            \\ 
\textbf{Data Set Type}                & \textbf{Train} & \textbf{Test}                             &                                           & \textbf{PR}                               & \textbf{ROC}                              & \textbf{Train}        & \textbf{Test}                             &                                           & \textbf{PR}                               & \textbf{ROC}                               \\ 
\hline
Without Synthetic                     & 0.734          & {\cellcolor[rgb]{0.388,0.745,0.482}}0.649 & {\cellcolor[rgb]{0.753,0.902,0.961}}0.205 & {\cellcolor[rgb]{0.949,0.808,0.937}}0.313 & {\cellcolor[rgb]{0.949,0.949,0.949}}0.524 &                       &                                           &                                           &                                           &                                            \\ 
\hline
30\% Minority with Synthetic           & 0.733          & {\cellcolor[rgb]{0.455,0.773,0.541}}0.645 & {\cellcolor[rgb]{0.718,0.89,0.957}}0.239  & {\cellcolor[rgb]{0.945,0.792,0.933}}0.332 & {\cellcolor[rgb]{0.949,0.949,0.918}}0.530 & 0.733                 & {\cellcolor[rgb]{0.388,0.745,0.482}}0.654 & {\cellcolor[rgb]{0.753,0.902,0.961}}0.271 & {\cellcolor[rgb]{0.949,0.808,0.937}}0.376 & {\cellcolor[rgb]{0.949,0.949,0.949}}0.553  \\ 
\hline
32\% Minority with Synthetic           & 0.723          & {\cellcolor[rgb]{0.863,0.937,0.894}}0.624 & {\cellcolor[rgb]{0.627,0.851,0.941}}0.322 & {\cellcolor[rgb]{0.929,0.729,0.91}}0.401  & {\cellcolor[rgb]{0.957,0.957,0.741}}0.563 & 0.728                 & {\cellcolor[rgb]{0.749,0.894,0.796}}0.636 & {\cellcolor[rgb]{0.682,0.875,0.949}}0.328 & {\cellcolor[rgb]{0.941,0.769,0.925}}0.414 & {\cellcolor[rgb]{0.953,0.953,0.839}}0.571  \\ 
\hline
34\% Minority with Synthetic           & 0.724          & {\cellcolor[rgb]{0.824,0.922,0.859}}0.626 & {\cellcolor[rgb]{0.588,0.835,0.937}}0.357 & {\cellcolor[rgb]{0.922,0.694,0.898}}0.440 & {\cellcolor[rgb]{0.961,0.961,0.675}}0.576 & 0.734                 & {\cellcolor[rgb]{0.675,0.863,0.729}}0.640 & {\cellcolor[rgb]{0.62,0.851,0.941}}0.375  & {\cellcolor[rgb]{0.922,0.706,0.902}}0.474 & {\cellcolor[rgb]{0.961,0.961,0.714}}0.593  \\ 
\hline
36\% Minority with Synthetic           & 0.721          & {\cellcolor[rgb]{0.984,0.988,0.996}}0.618 & {\cellcolor[rgb]{0.541,0.816,0.929}}0.401 & {\cellcolor[rgb]{0.906,0.647,0.882}}0.491 & {\cellcolor[rgb]{0.969,0.969,0.557}}0.598 & 0.736                 & {\cellcolor[rgb]{0.824,0.922,0.859}}0.632 & {\cellcolor[rgb]{0.573,0.827,0.933}}0.414 & {\cellcolor[rgb]{0.91,0.655,0.886}}0.524  & {\cellcolor[rgb]{0.965,0.965,0.584}}0.614  \\ 
\hline
38\% Minority with Synthetic           & 0.727          & {\cellcolor[rgb]{0.867,0.941,0.898}}0.624 & {\cellcolor[rgb]{0.486,0.796,0.922}}0.449 & {\cellcolor[rgb]{0.894,0.604,0.867}}0.540 & {\cellcolor[rgb]{0.973,0.973,0.443}}0.620 & 0.727                 & {\cellcolor[rgb]{0.988,0.988,1}}0.624     & {\cellcolor[rgb]{0.514,0.804,0.925}}0.459 & {\cellcolor[rgb]{0.902,0.631,0.875}}0.547 & {\cellcolor[rgb]{0.969,0.969,0.529}}0.623  \\ 
\hline
40\% Minority with Synthetic           & 0.725          & {\cellcolor[rgb]{0.988,0.988,1}}0.618     & {\cellcolor[rgb]{0.431,0.773,0.91}}0.501  & {\cellcolor[rgb]{0.886,0.569,0.855}}0.575 & {\cellcolor[rgb]{0.98,0.98,0.361}}0.635   & 0.734                 & {\cellcolor[rgb]{0.847,0.933,0.878}}0.631 & {\cellcolor[rgb]{0.455,0.78,0.914}}0.507  & {\cellcolor[rgb]{0.89,0.584,0.859}}0.591  & {\cellcolor[rgb]{0.976,0.976,0.424}}0.641  \\ 
\hline
42\% Minority with Synthetic           & 0.733          & {\cellcolor[rgb]{0.82,0.922,0.855}}0.626  & {\cellcolor[rgb]{0.396,0.757,0.906}}0.533 & {\cellcolor[rgb]{0.878,0.537,0.843}}0.611 & {\cellcolor[rgb]{0.98,0.98,0.294}}0.648   & 0.736                 & {\cellcolor[rgb]{0.722,0.882,0.769}}0.638 & {\cellcolor[rgb]{0.404,0.761,0.906}}0.546 & {\cellcolor[rgb]{0.878,0.537,0.843}}0.634 & {\cellcolor[rgb]{0.98,0.98,0.314}}0.660    \\ 
\hline
45\% Minority with Synthetic           & 0.734          & {\cellcolor[rgb]{0.733,0.886,0.78}}0.631  & {\cellcolor[rgb]{0.349,0.737,0.898}}0.575 & {\cellcolor[rgb]{0.867,0.498,0.831}}0.656 & {\cellcolor[rgb]{0.988,0.988,0.204}}0.665 & 0.736                 & {\cellcolor[rgb]{0.718,0.878,0.769}}0.638 & {\cellcolor[rgb]{0.353,0.737,0.898}}0.586 & {\cellcolor[rgb]{0.867,0.494,0.827}}0.676 & {\cellcolor[rgb]{0.984,0.984,0.224}}0.675  \\ 
\hline
48\% Minority with Synthetic           & 0.749          & {\cellcolor[rgb]{0.537,0.808,0.612}}0.641 & {\cellcolor[rgb]{0.302,0.718,0.89}}0.619  & {\cellcolor[rgb]{0.855,0.455,0.816}}0.700 & {\cellcolor[rgb]{0.992,0.992,0.09}}0.686  & 0.742                 & {\cellcolor[rgb]{0.631,0.847,0.694}}0.643 & {\cellcolor[rgb]{0.294,0.714,0.89}}0.633  & {\cellcolor[rgb]{0.855,0.451,0.812}}0.718 & {\cellcolor[rgb]{0.992,0.992,0.102}}0.695  \\ 
\hline
50\% Minority with Synthetic           & 0.752          & {\cellcolor[rgb]{0.392,0.749,0.486}}0.649 & {\cellcolor[rgb]{0.267,0.702,0.882}}0.650 & {\cellcolor[rgb]{0.847,0.427,0.804}}0.730 & {\cellcolor{yellow}}0.703                 & 0.756                 & {\cellcolor[rgb]{0.388,0.745,0.482}}0.655 & {\cellcolor[rgb]{0.267,0.702,0.882}}0.652 & {\cellcolor[rgb]{0.847,0.427,0.804}}0.737 & {\cellcolor{yellow}}0.712                  \\ 
\hline
\multicolumn{11}{|c|}{{\cellcolor[rgb]{0.753,0.624,0.984}}\textbf{NN Classifier}}                                                                                                                                                                                                                                            \\ 
\hline
\multicolumn{1}{|c|}{\textbf{Scores}} & \multicolumn{2}{c|}{\textbf{Accuracy Score}}               & \multirow{2}{*}{\textbf{F1 Score}}        & \multicolumn{2}{c|}{\textbf{AUC Score}}                                               & \multicolumn{2}{c|}{\textbf{Accuracy Score AOL}}                  & \multirow{2}{*}{\textbf{F1 Score AOL}}    & \multicolumn{2}{c|}{\textbf{AUC Score AOL}}                                            \\ 
\textbf{Data Set Type}                & \textbf{Train} & \textbf{Test}                             &                                           & \textbf{PR}                               & \textbf{ROC}                              & \textbf{Train}        & \textbf{Test}                             &                                           & \textbf{PR}                               & \textbf{ROC}                               \\ 
\hline
Without Synthetic                     & 0.716          & {\cellcolor[rgb]{0.733,0.886,0.78}}0.709  & {\cellcolor[rgb]{0.753,0.902,0.961}}0.041 & {\cellcolor[rgb]{0.949,0.808,0.937}}0.368 & {\cellcolor[rgb]{0.949,0.949,0.949}}0.576 &                       &                                           &                                           &                                           &                                            \\ 
\hline
30\% Minority with Synthetic           & 0.703          & {\cellcolor[rgb]{0.757,0.894,0.8}}0.703   & {\cellcolor[rgb]{0.749,0.902,0.961}}0.048 & {\cellcolor[rgb]{0.949,0.8,0.933}}0.382   & {\cellcolor[rgb]{0.949,0.949,0.925}}0.583 & 0.706                 & {\cellcolor[rgb]{0.388,0.745,0.482}}0.704 & {\cellcolor[rgb]{0.733,0.894,0.961}}0.081 & {\cellcolor[rgb]{0.949,0.808,0.937}}0.420 & {\cellcolor[rgb]{0.949,0.949,0.949}}0.594  \\ 
\hline
32\% Minority with Synthetic           & 0.688          & {\cellcolor[rgb]{0.855,0.937,0.886}}0.678 & {\cellcolor[rgb]{0.733,0.894,0.961}}0.070 & {\cellcolor[rgb]{0.941,0.773,0.925}}0.416 & {\cellcolor[rgb]{0.949,0.949,0.902}}0.588 & 0.686                 & {\cellcolor[rgb]{0.831,0.925,0.867}}0.683 & {\cellcolor[rgb]{0.753,0.902,0.961}}0.052 & {\cellcolor[rgb]{0.949,0.8,0.937}}0.432   & {\cellcolor[rgb]{0.949,0.949,0.937}}0.597  \\ 
\hline
34\% Minority with Synthetic           & 0.666          & {\cellcolor[rgb]{0.922,0.961,0.945}}0.661 & {\cellcolor[rgb]{0.741,0.898,0.961}}0.059 & {\cellcolor[rgb]{0.941,0.769,0.925}}0.423 & {\cellcolor[rgb]{0.949,0.949,0.925}}0.583 & 0.671                 & {\cellcolor[rgb]{0.906,0.957,0.929}}0.665 & {\cellcolor[rgb]{0.714,0.886,0.957}}0.112 & {\cellcolor[rgb]{0.941,0.78,0.929}}0.457  & {\cellcolor[rgb]{0.949,0.949,0.933}}0.598  \\ 
\hline
36\% Minority with Synthetic           & 0.647          & {\cellcolor[rgb]{0.984,0.988,1}}0.645     & {\cellcolor[rgb]{0.71,0.886,0.957}}0.102  & {\cellcolor[rgb]{0.933,0.749,0.918}}0.446 & {\cellcolor[rgb]{0.949,0.949,0.898}}0.589 & 0.656                 & {\cellcolor[rgb]{0.988,0.988,1}}0.644     & {\cellcolor[rgb]{0.659,0.867,0.949}}0.187 & {\cellcolor[rgb]{0.937,0.753,0.922}}0.486 & {\cellcolor[rgb]{0.949,0.949,0.949}}0.594  \\ 
\hline
38\% Minority with Synthetic           & 0.740          & {\cellcolor[rgb]{0.608,0.835,0.675}}0.740 & {\cellcolor[rgb]{0.427,0.769,0.91}}0.497  & {\cellcolor[rgb]{0.882,0.553,0.851}}0.706 & {\cellcolor[rgb]{0.976,0.976,0.392}}0.718 & 0.749                 & {\cellcolor[rgb]{0.608,0.835,0.675}}0.740 & {\cellcolor[rgb]{0.439,0.773,0.91}}0.506  & {\cellcolor[rgb]{0.886,0.569,0.855}}0.706 & {\cellcolor[rgb]{0.973,0.973,0.455}}0.714  \\ 
\hline
40\% Minority with Synthetic           & 0.751          & {\cellcolor[rgb]{0.588,0.827,0.655}}0.745 & {\cellcolor[rgb]{0.388,0.753,0.902}}0.553 & {\cellcolor[rgb]{0.875,0.518,0.835}}0.751 & {\cellcolor[rgb]{0.98,0.98,0.302}}0.741   & 0.761                 & {\cellcolor[rgb]{0.565,0.816,0.635}}0.751 & {\cellcolor[rgb]{0.388,0.753,0.902}}0.577 & {\cellcolor[rgb]{0.875,0.525,0.839}}0.762 & {\cellcolor[rgb]{0.98,0.98,0.302}}0.750    \\ 
\hline
42\% Minority with Synthetic           & 0.759          & {\cellcolor[rgb]{0.537,0.808,0.612}}0.758 & {\cellcolor[rgb]{0.341,0.733,0.898}}0.616 & {\cellcolor[rgb]{0.867,0.498,0.827}}0.780 & {\cellcolor[rgb]{0.984,0.984,0.235}}0.758 & 0.759                 & {\cellcolor[rgb]{0.533,0.804,0.608}}0.759 & {\cellcolor[rgb]{0.361,0.741,0.898}}0.615 & {\cellcolor[rgb]{0.871,0.502,0.831}}0.786 & {\cellcolor[rgb]{0.984,0.984,0.255}}0.762  \\ 
\hline
45\% Minority with Synthetic           & 0.767          & {\cellcolor[rgb]{0.49,0.788,0.573}}0.770  & {\cellcolor[rgb]{0.31,0.722,0.89}}0.664   & {\cellcolor[rgb]{0.859,0.467,0.82}}0.820  & {\cellcolor[rgb]{0.992,0.992,0.133}}0.783 & 0.772                 & {\cellcolor[rgb]{0.51,0.796,0.588}}0.765  & {\cellcolor[rgb]{0.322,0.725,0.894}}0.671 & {\cellcolor[rgb]{0.859,0.471,0.82}}0.824  & {\cellcolor[rgb]{0.988,0.988,0.173}}0.782  \\ 
\hline
48\% Minority with Synthetic           & 0.780          & {\cellcolor[rgb]{0.471,0.78,0.553}}0.775  & {\cellcolor[rgb]{0.278,0.71,0.886}}0.705  & {\cellcolor[rgb]{0.851,0.443,0.812}}0.850 & {\cellcolor[rgb]{0.996,0.996,0.067}}0.800 & 0.780                 & {\cellcolor[rgb]{0.451,0.773,0.537}}0.780 & {\cellcolor[rgb]{0.286,0.714,0.886}}0.719 & {\cellcolor[rgb]{0.851,0.443,0.812}}0.861 & {\cellcolor[rgb]{0.996,0.996,0.051}}0.810  \\ 
\hline
50\% Minority with Synthetic           & 0.777          & {\cellcolor[rgb]{0.463,0.776,0.545}}0.777 & {\cellcolor[rgb]{0.267,0.702,0.882}}0.719 & {\cellcolor[rgb]{0.847,0.427,0.804}}0.868 & {\cellcolor{yellow}}0.817                 & 0.794                 & {\cellcolor[rgb]{0.388,0.745,0.482}}0.795 & {\cellcolor[rgb]{0.267,0.702,0.882}}0.747 & {\cellcolor[rgb]{0.847,0.427,0.804}}0.875 & {\cellcolor{yellow}}0.822                  \\ 
\hline
\end{tabular}
}
\caption{Table detailing Model evaluation parameters across RF, KNN, and NN as SMOTE \% is increased from 30\% to 50\% each step including outlier boosting and its impacts.}
\label{Model_Evaluation_Stats}
\end{table}

\section{Improvements Due to AOL}
As we observe the changes in classification stats due to Angular Outlier boosting, we can summarize the improvements under the following heads. Table \ref{Improvemnt Table} quantifies the improvements for all statistics across three models. While the previous section assesses the overall model performance, in this section, we outline the improvement trends in the model stats.

\subsection{Train Accuracy Improvement}

For \textbf{RF}, the boost in Train Accuracy is modest, with the best improvement being just \textbf{0.18\%} at 34\% and 36\% synthetic data. This suggests that while Angular Outlier Boost (AOL) does help, its impact on RF is minimal. On the other hand, \textbf{KNN} shows a more noticeable improvement, peaking at \textbf{2.04\%} with 36\% synthetic data, making it clear that KNN benefits significantly from AOL, particularly at moderate synthetic levels. \textbf{NN} outperform both RF and KNN for this metric, with a maximum gain of \textbf{2.20\%} at 50\% synthetic data. NN thrives with AOL as synthetic data levels increase, showcasing its adaptability to this boosting technique.

\begin{figure}[H]
    \centering
    \includegraphics[width=1.2\linewidth]{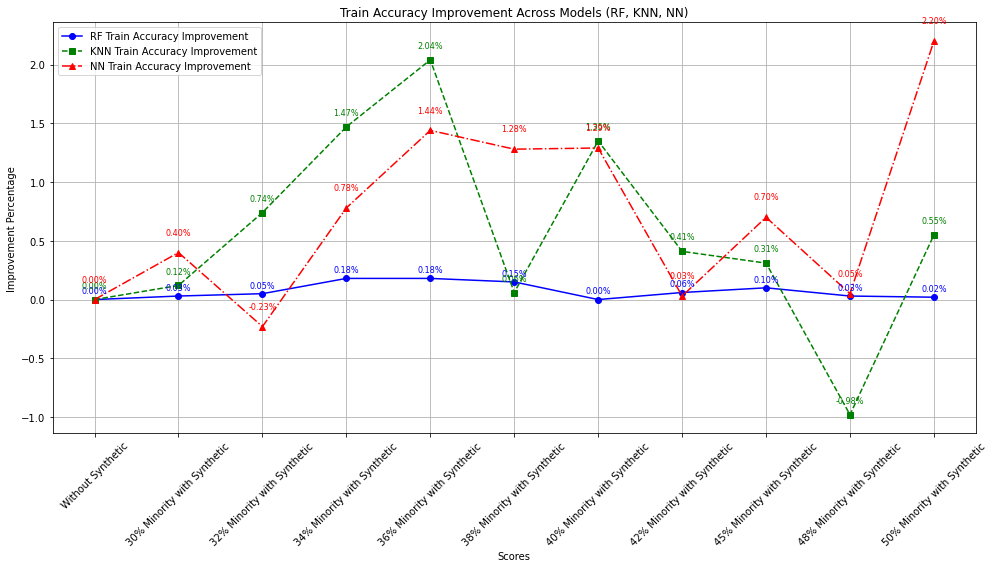}
    \caption{Figure Showing Improvement trend of Train accuracy due to AOL boost  across RF, NN, KNN with varying levels of Quantum-SMOTEV2}
    \label{Train_Accuracy_Improvement}
\end{figure}

\subsection{Test Accuracy Improvement}

When it comes to Test Accuracy, \textbf{RF} achieves its best improvement of \textbf{1.72\%} at 32\% synthetic data but struggles at higher levels, even experiencing declines. This shows that RF’s benefits from AOL are limited to specific configurations. \textbf{KNN}, however, performs much better, reaching a \textbf{2.36\%} improvement at 36\% synthetic data and maintaining consistent gains, highlighting its compatibility with AOL. \textbf{NN} shows mixed results; while it achieves a high of \textbf{2.30\%} at 50\% synthetic data, some configurations (like 38\% synthetic) lead to slight performance drops. This reflects NN’s sensitivity to how AOL is applied, requiring careful tuning.

\begin{figure}[H]
    \centering
    \includegraphics[width=1.2\linewidth]{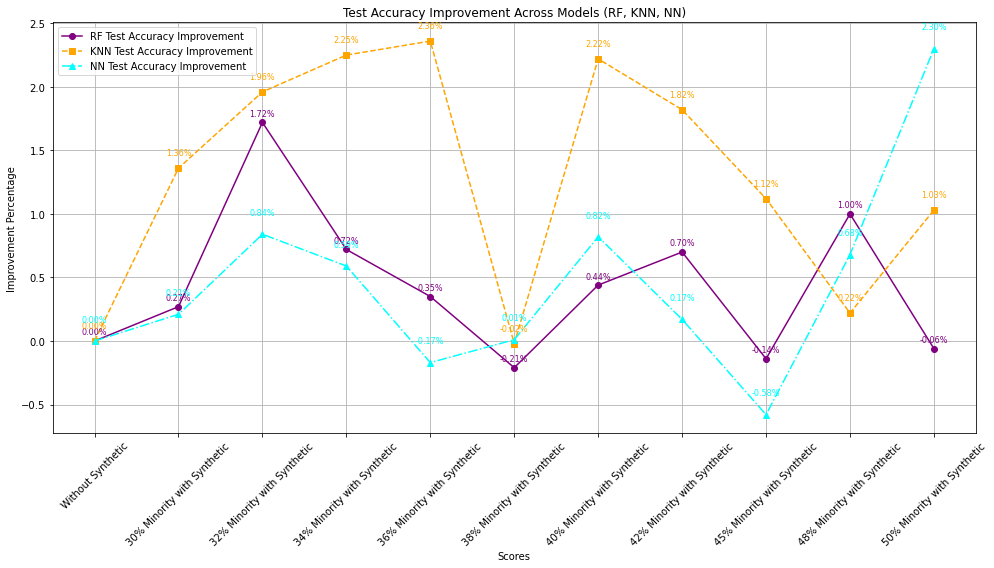}
    \caption{Figure Showing Improvement trend of Test accuracy due to AOL boost  across RF, NN, KNN with varying levels of Quantum-SMOTEV2}
    \label{Test_Accuracy_Improvement}
\end{figure}

\subsection{F1 Score Improvement}

In terms of F1 Score, \textbf{RF} benefits significantly at lower synthetic levels, with a peak improvement of \textbf{13.00\%} at 30\% synthetic data. However, the benefits diminish quickly as synthetic levels increase. \textbf{KNN} follows a similar pattern, achieving a maximum gain of \textbf{13.10\%} at 30\% synthetic data and maintaining slightly better performance than RF as synthetic levels rise. \textbf{NN}, however, is the clear winner for F1 Score. It sees an extraordinary \textbf{91.28\%} improvement at 34\% synthetic data and \textbf{84.45\%} at 36\%, showing that AOL is incredibly effective for NN in balancing precision and recall, especially at mid-range synthetic levels.

\begin{figure}[H]
    \centering
    \includegraphics[width=1.2\linewidth]{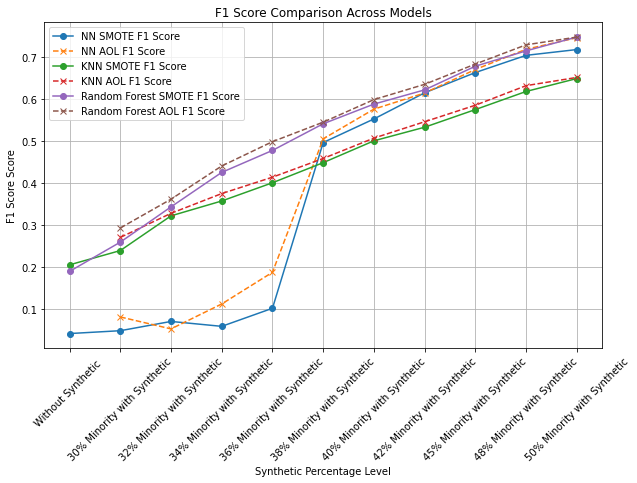}
    \caption{Figure Showing Improvement trend of F1 Score due to AOL boost  across RF, NN, KNN with varying levels of Quantum-SMOTEV2}
    \label{F1_Score_Improvement}
\end{figure}

\subsection{PR Score Improvement}

For PR Score, \textbf{RF} shows a moderate peak improvement of \textbf{8.07\%} at 30\% synthetic data, but its gains quickly taper off as synthetic levels increase. \textbf{KNN} performs much better, achieving a significant boost of \textbf{13.23\%} at 30\% synthetic data and maintaining strong gains across configurations. \textbf{NN} also sees notable improvements, with a peak of \textbf{9.89\%} at 30\% synthetic data, though its benefits are less consistent at higher levels, with some configurations offering negligible gains.

\begin{figure}[H]
    \centering
    \includegraphics[width=1.2\linewidth]{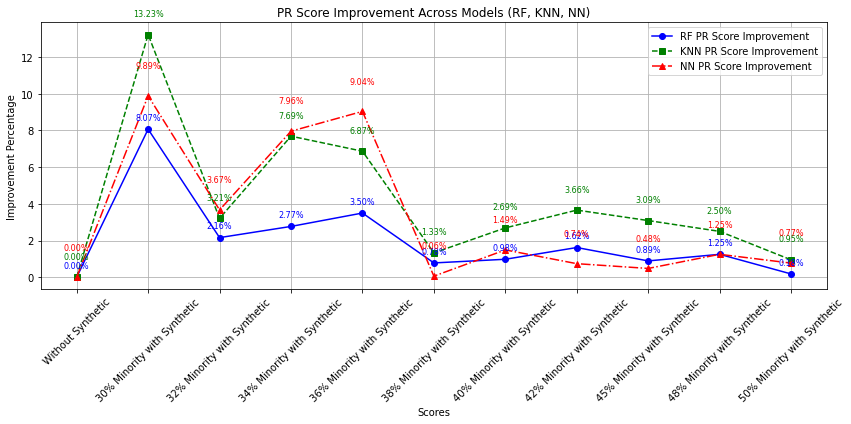}
    \caption{Figure Showing Improvement trend of Precision-Recall Score due to AOL boost  across RF, NN, KNN with varying levels of Quantum-SMOTEV2}
    \label{PR_Score_Improvement}
\end{figure}

\subsection{ROC/AUC Score Improvement}

\textbf{RF} sees moderate improvements in ROC/AUC, with the best result being \textbf{3.59\%} at 30\% synthetic data. However, it fails to maintain momentum at higher synthetic levels. \textbf{KNN} shines in this metric, with a peak improvement of \textbf{4.29\%} at 30\% synthetic data, demonstrating its robustness when combined with AOL. \textbf{NN}, on the other hand, shows moderate gains, peaking at \textbf{2.52\%} at 34\% synthetic data. However, its performance drops with certain configurations, such as a \textbf{-0.59\%} decline at 38\% synthetic data, reflecting its sensitivity to AOL.

\begin{figure}[H]
    \centering
    \includegraphics[width=1.2\linewidth]{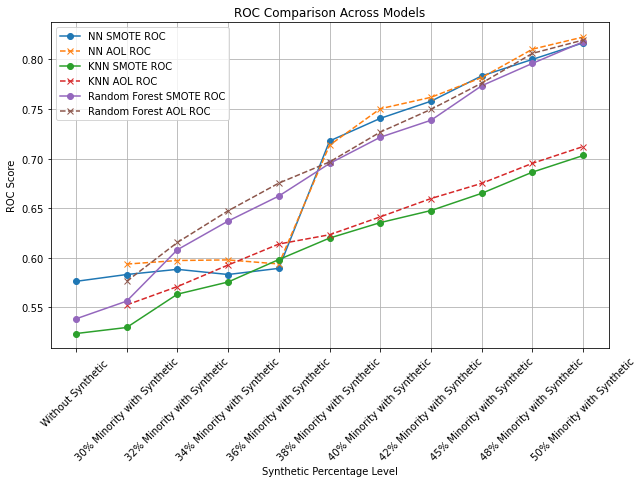}
    \caption{Figure Showing Improvement trend of ROC curve due to AOL boost  across RF, NN, KNN with varying levels of Quantum-SMOTEV2}
    \label{ROC_AUC_Improvement}
\end{figure}

\subsection{Overall Observations}

Angular Outlier Boost (AOL) significantly enhances classification performance, but its effectiveness varies across models and metrics. \textbf{NN} emerge as the top performer for most metrics, particularly F1 Score and Train Accuracy, where the gains are substantial, especially at mid-to-high synthetic levels. \textbf{KNN} also performs exceptionally well, with consistent and noticeable improvements across all metrics, making it highly adaptable to AOL. \textbf{RF}, while benefiting modestly, shows limited and inconsistent gains, particularly at higher synthetic levels.

Moderate synthetic levels (30–36\%) seem to be the sweet spot for AOL, providing the best balance between boosting performance and maintaining model stability. Overall, AOL proves to be a powerful technique for enhancing classification performance, with \textbf{NN and KNN} being the most responsive to its benefits. This makes them ideal choices for scenarios where AOL can be leveraged to its full potential.

\begin{table}[hbt!]
\centering
\resizebox{\columnwidth}{!}{%
\begin{tabular}{|lccccc|}
\hline
\multicolumn{6}{|c|}{\cellcolor[HTML]{CAEDFB}} \\
\multicolumn{6}{|c|}{\multirow{-2}{*}{\cellcolor[HTML]{CAEDFB}\textbf{RF  Improvement Statistics}}} \\ \hline
\multicolumn{1}{|c|}{\textbf{Scores}} &
  \multicolumn{1}{l|}{\textbf{Train Accuracy}} &
  \multicolumn{1}{l|}{\textbf{Test Accuracy}} &
  \multicolumn{1}{l|}{\textbf{F1 Score}} &
  \multicolumn{1}{l|}{\textbf{PR Score}} &
  \multicolumn{1}{l|}{\textbf{ROC/AUC Score}} \\ \hline
\multicolumn{1}{|l|}{Without   Synthetic} &
  \multicolumn{1}{c|}{\cellcolor[HTML]{FDFEFF}0.00\%} &
  \multicolumn{1}{c|}{\cellcolor[HTML]{ECFBEE}0.00\%} &
  \multicolumn{1}{c|}{\cellcolor[HTML]{FFFFFF}0.00\%} &
  \multicolumn{1}{c|}{\cellcolor[HTML]{FFFFFF}0.00\%} &
  \cellcolor[HTML]{FFFFFF}0.00\% \\ \hline
\multicolumn{1}{|l|}{30\% Minority   with Synthetic} &
  \multicolumn{1}{c|}{\cellcolor[HTML]{E2EEF9}0.03\%} &
  \multicolumn{1}{c|}{\cellcolor[HTML]{D3F5D7}0.27\%} &
  \multicolumn{1}{c|}{\cellcolor[HTML]{F1A983}13.00\%} &
  \multicolumn{1}{c|}{\cellcolor[HTML]{D86DCD}8.07\%} &
  \cellcolor[HTML]{12E7F2}3.59\% \\ \hline
\multicolumn{1}{|l|}{32\% Minority   with Synthetic} &
  \multicolumn{1}{c|}{\cellcolor[HTML]{CBDFF4}0.05\%} &
  \multicolumn{1}{c|}{\cellcolor[HTML]{47D359}1.72\%} &
  \multicolumn{1}{c|}{\cellcolor[HTML]{FADCCD}5.33\%} &
  \multicolumn{1}{c|}{\cellcolor[HTML]{F5D8F2}2.16\%} &
  \cellcolor[HTML]{B0F7FB}1.21\% \\ \hline
\multicolumn{1}{|l|}{34\% Minority   with Synthetic} &
  \multicolumn{1}{c|}{\cellcolor[HTML]{4D93D9}0.18\%} &
  \multicolumn{1}{c|}{\cellcolor[HTML]{A7EAB0}0.72\%} &
  \multicolumn{1}{c|}{\cellcolor[HTML]{FCE8DD}3.59\%} &
  \multicolumn{1}{c|}{\cellcolor[HTML]{F2CDEE}2.77\%} &
  \cellcolor[HTML]{95F5FA}1.62\% \\ \hline
\multicolumn{1}{|l|}{36\% Minority   with Synthetic} &
  \multicolumn{1}{c|}{\cellcolor[HTML]{5397DB}0.18\%} &
  \multicolumn{1}{c|}{\cellcolor[HTML]{CAF3CF}0.35\%} &
  \multicolumn{1}{c|}{\cellcolor[HTML]{FBE3D6}4.36\%} &
  \multicolumn{1}{c|}{\cellcolor[HTML]{EFC0EA}3.50\%} &
  \cellcolor[HTML]{7EF2F8}1.97\% \\ \hline
\multicolumn{1}{|l|}{38\% Minority   with Synthetic} &
  \multicolumn{1}{c|}{\cellcolor[HTML]{6BA6E0}0.15\%} &
  \multicolumn{1}{c|}{\cellcolor[HTML]{FFFFFF}-0.21\%} &
  \multicolumn{1}{c|}{\cellcolor[HTML]{FFFBF9}0.70\%} &
  \multicolumn{1}{c|}{\cellcolor[HTML]{FCF1FB}0.78\%} &
  \cellcolor[HTML]{F1FEFF}0.22\% \\ \hline
\multicolumn{1}{|l|}{40\% Minority   with Synthetic} &
  \multicolumn{1}{c|}{\cellcolor[HTML]{FFFFFF}0.00\%} &
  \multicolumn{1}{c|}{\cellcolor[HTML]{C2F1C8}0.44\%} &
  \multicolumn{1}{c|}{\cellcolor[HTML]{FEF3EE}1.82\%} &
  \multicolumn{1}{c|}{\cellcolor[HTML]{FBEEF9}0.98\%} &
  \cellcolor[HTML]{D0FBFD}0.72\% \\ \hline
\multicolumn{1}{|l|}{42\% Minority   with Synthetic} &
  \multicolumn{1}{c|}{\cellcolor[HTML]{C7DDF3}0.06\%} &
  \multicolumn{1}{c|}{\cellcolor[HTML]{A9EBB1}0.70\%} &
  \multicolumn{1}{c|}{\cellcolor[HTML]{FDF1EB}2.12\%} &
  \multicolumn{1}{c|}{\cellcolor[HTML]{F8E2F5}1.62\%} &
  \cellcolor[HTML]{9DF6FA}1.49\% \\ \hline
\multicolumn{1}{|l|}{45\% Minority   with Synthetic} &
  \multicolumn{1}{c|}{\cellcolor[HTML]{99C1EA}0.10\%} &
  \multicolumn{1}{c|}{\cellcolor[HTML]{FAFEFA}-0.14\%} &
  \multicolumn{1}{c|}{\cellcolor[HTML]{FFFAF8}0.76\%} &
  \multicolumn{1}{c|}{\cellcolor[HTML]{FBEFFA}0.89\%} &
  \cellcolor[HTML]{E9FDFE}0.34\% \\ \hline
\multicolumn{1}{|l|}{48\% Minority   with Synthetic} &
  \multicolumn{1}{c|}{\cellcolor[HTML]{DDEAF8}0.03\%} &
  \multicolumn{1}{c|}{\cellcolor[HTML]{8CE498}1.00\%} &
  \multicolumn{1}{c|}{\cellcolor[HTML]{FDF2EC}2.01\%} &
  \multicolumn{1}{c|}{\cellcolor[HTML]{F9E9F8}1.25\%} &
  \cellcolor[HTML]{ACF7FB}1.27\% \\ \hline
\multicolumn{1}{|l|}{50\% Minority   with Synthetic} &
  \multicolumn{1}{c|}{\cellcolor[HTML]{E6F0FA}0.02\%} &
  \multicolumn{1}{c|}{\cellcolor[HTML]{F1FCF3}-0.06\%} &
  \multicolumn{1}{c|}{\cellcolor[HTML]{FFFFFF}-0.01\%} &
  \multicolumn{1}{c|}{\cellcolor[HTML]{FFFCFE}0.18\%} &
  \cellcolor[HTML]{F1FEFF}0.22\% \\ \hline
\multicolumn{6}{|c|}{\cellcolor[HTML]{DAF2D0}} \\
\multicolumn{6}{|c|}{\multirow{-2}{*}{\cellcolor[HTML]{DAF2D0}\textbf{KNN Improvement Statistics}}} \\ \hline
\multicolumn{1}{|c|}{\textbf{Scores}} &
  \multicolumn{1}{l|}{\textbf{Train Accuracy}} &
  \multicolumn{1}{l|}{\textbf{Test Accuracy}} &
  \multicolumn{1}{l|}{\textbf{F1 Score}} &
  \multicolumn{1}{l|}{\textbf{PR Score}} &
  \multicolumn{1}{l|}{\textbf{ROC/AUC Score}} \\ \hline
\multicolumn{1}{|l|}{Without   Synthetic} &
  \multicolumn{1}{c|}{\cellcolor[HTML]{C6DCF3}0.00\%} &
  \multicolumn{1}{c|}{\cellcolor[HTML]{FEFFFE}0.00\%} &
  \multicolumn{1}{c|}{\cellcolor[HTML]{FFFFFF}0.00\%} &
  \multicolumn{1}{c|}{\cellcolor[HTML]{FFFFFF}0.00\%} &
  \cellcolor[HTML]{FFFFFF}0.00\% \\ \hline
\multicolumn{1}{|l|}{30\% Minority   with Synthetic} &
  \multicolumn{1}{c|}{\cellcolor[HTML]{BFD8F2}0.12\%} &
  \multicolumn{1}{c|}{\cellcolor[HTML]{95E69F}1.36\%} &
  \multicolumn{1}{c|}{\cellcolor[HTML]{F1A983}13.10\%} &
  \multicolumn{1}{c|}{\cellcolor[HTML]{D86DCD}13.23\%} &
  \cellcolor[HTML]{12E7F2}4.29\% \\ \hline
\multicolumn{1}{|l|}{32\% Minority   with Synthetic} &
  \multicolumn{1}{c|}{\cellcolor[HTML]{9AC2EA}0.74\%} &
  \multicolumn{1}{c|}{\cellcolor[HTML]{67DB76}1.96\%} &
  \multicolumn{1}{c|}{\cellcolor[HTML]{FDF3EE}1.87\%} &
  \multicolumn{1}{c|}{\cellcolor[HTML]{F6DCF3}3.21\%} &
  \cellcolor[HTML]{B5F8FB}1.36\% \\ \hline
\multicolumn{1}{|l|}{34\% Minority   with Synthetic} &
  \multicolumn{1}{c|}{\cellcolor[HTML]{6FA8E1}1.47\%} &
  \multicolumn{1}{c|}{\cellcolor[HTML]{50D661}2.25\%} &
  \multicolumn{1}{c|}{\cellcolor[HTML]{FADFD1}4.95\%} &
  \multicolumn{1}{c|}{\cellcolor[HTML]{E9ABE2}7.69\%} &
  \cellcolor[HTML]{5BEFF7}2.97\% \\ \hline
\multicolumn{1}{|l|}{36\% Minority   with Synthetic} &
  \multicolumn{1}{c|}{\cellcolor[HTML]{4D93D9}2.04\%} &
  \multicolumn{1}{c|}{\cellcolor[HTML]{47D359}2.36\%} &
  \multicolumn{1}{c|}{\cellcolor[HTML]{FCEAE0}3.31\%} &
  \multicolumn{1}{c|}{\cellcolor[HTML]{EBB4E6}6.87\%} &
  \cellcolor[HTML]{6FF1F8}2.61\% \\ \hline
\multicolumn{1}{|l|}{38\% Minority   with Synthetic} &
  \multicolumn{1}{c|}{\cellcolor[HTML]{C2DAF2}0.06\%} &
  \multicolumn{1}{c|}{\cellcolor[HTML]{FFFFFF}-0.02\%} &
  \multicolumn{1}{c|}{\cellcolor[HTML]{FDF1EA}2.23\%} &
  \multicolumn{1}{c|}{\cellcolor[HTML]{FCF1FA}1.33\%} &
  \cellcolor[HTML]{E2FCFE}0.54\% \\ \hline
\multicolumn{1}{|l|}{40\% Minority   with Synthetic} &
  \multicolumn{1}{c|}{\cellcolor[HTML]{76ACE2}1.35\%} &
  \multicolumn{1}{c|}{\cellcolor[HTML]{53D664}2.22\%} &
  \multicolumn{1}{c|}{\cellcolor[HTML]{FEF7F3}1.29\%} &
  \multicolumn{1}{c|}{\cellcolor[HTML]{F8E2F5}2.69\%} &
  \cellcolor[HTML]{CDFAFD}0.91\% \\ \hline
\multicolumn{1}{|l|}{42\% Minority   with Synthetic} &
  \multicolumn{1}{c|}{\cellcolor[HTML]{AECEEE}0.41\%} &
  \multicolumn{1}{c|}{\cellcolor[HTML]{72DE80}1.82\%} &
  \multicolumn{1}{c|}{\cellcolor[HTML]{FDEFE8}2.45\%} &
  \multicolumn{1}{c|}{\cellcolor[HTML]{F5D7F2}3.66\%} &
  \cellcolor[HTML]{98F5FA}1.87\% \\ \hline
\multicolumn{1}{|l|}{45\% Minority   with Synthetic} &
  \multicolumn{1}{c|}{\cellcolor[HTML]{B3D1EF}0.31\%} &
  \multicolumn{1}{c|}{\cellcolor[HTML]{A7EAB0}1.12\%} &
  \multicolumn{1}{c|}{\cellcolor[HTML]{FEF4EE}1.81\%} &
  \multicolumn{1}{c|}{\cellcolor[HTML]{F6DDF4}3.09\%} &
  \cellcolor[HTML]{ACF7FB}1.51\% \\ \hline
\multicolumn{1}{|l|}{48\% Minority   with Synthetic} &
  \multicolumn{1}{c|}{\cellcolor[HTML]{FFFFFF}-0.98\%} &
  \multicolumn{1}{c|}{\cellcolor[HTML]{EDFBEE}0.22\%} &
  \multicolumn{1}{c|}{\cellcolor[HTML]{FDF1EA}2.24\%} &
  \multicolumn{1}{c|}{\cellcolor[HTML]{F8E4F6}2.50\%} &
  \cellcolor[HTML]{B8F8FC}1.30\% \\ \hline
\multicolumn{1}{|l|}{50\% Minority   with Synthetic} &
  \multicolumn{1}{c|}{\cellcolor[HTML]{A5C9EC}0.55\%} &
  \multicolumn{1}{c|}{\cellcolor[HTML]{AEECB6}1.03\%} &
  \multicolumn{1}{c|}{\cellcolor[HTML]{FFFDFC}0.40\%} &
  \multicolumn{1}{c|}{\cellcolor[HTML]{FDF5FC}0.95\%} &
  \cellcolor[HTML]{B9F8FC}1.29\% \\ \hline
\multicolumn{6}{|c|}{\cellcolor[HTML]{C09FFB}} \\
\multicolumn{6}{|c|}{\multirow{-2}{*}{\cellcolor[HTML]{C09FFB}\textbf{NN Improvement Ststistics}}} \\ \hline
\multicolumn{1}{|c|}{\textbf{Scores}} &
  \multicolumn{1}{l|}{\textbf{Train Accuracy}} &
  \multicolumn{1}{l|}{\textbf{Test Accuracy}} &
  \multicolumn{1}{l|}{\textbf{F1 Score}} &
  \multicolumn{1}{l|}{\textbf{PR Score}} &
  \multicolumn{1}{l|}{\textbf{ROC/AUC Score}} \\ \hline
\multicolumn{1}{|l|}{Without   Synthetic} &
  \multicolumn{1}{c|}{\cellcolor[HTML]{EEF5FC}0.00\%} &
  \multicolumn{1}{c|}{\cellcolor[HTML]{DAF7DE}0.00\%} &
  \multicolumn{1}{c|}{\cellcolor[HTML]{FCEDE5}0.00\%} &
  \multicolumn{1}{c|}{\cellcolor[HTML]{FFFFFF}0.00\%} &
  \cellcolor[HTML]{D3FBFD}0.00\% \\ \hline
\multicolumn{1}{|l|}{30\% Minority   with Synthetic} &
  \multicolumn{1}{c|}{\cellcolor[HTML]{D1E4F6}0.40\%} &
  \multicolumn{1}{c|}{\cellcolor[HTML]{CDF3D2}0.21\%} &
  \multicolumn{1}{c|}{\cellcolor[HTML]{F4BB9C}68.19\%} &
  \multicolumn{1}{c|}{\cellcolor[HTML]{D86DCD}9.89\%} &
  \cellcolor[HTML]{49EDF6}1.80\% \\ \hline
\multicolumn{1}{|l|}{32\% Minority   with Synthetic} &
  \multicolumn{1}{c|}{\cellcolor[HTML]{FFFFFF}-0.23\%} &
  \multicolumn{1}{c|}{\cellcolor[HTML]{A5EAAE}0.84\%} &
  \multicolumn{1}{c|}{\cellcolor[HTML]{FFFFFF}-25.36\%} &
  \multicolumn{1}{c|}{\cellcolor[HTML]{F1C9ED}3.67\%} &
  \cellcolor[HTML]{5FEFF7}1.51\% \\ \hline
\multicolumn{1}{|l|}{34\% Minority   with Synthetic} &
  \multicolumn{1}{c|}{\cellcolor[HTML]{B5D3F0}0.78\%} &
  \multicolumn{1}{c|}{\cellcolor[HTML]{B5EEBC}0.59\%} &
  \multicolumn{1}{c|}{\cellcolor[HTML]{F1A983}91.28\%} &
  \multicolumn{1}{c|}{\cellcolor[HTML]{E08AD7}7.96\%} &
  \cellcolor[HTML]{12E7F2}2.52\% \\ \hline
\multicolumn{1}{|l|}{36\% Minority   with Synthetic} &
  \multicolumn{1}{c|}{\cellcolor[HTML]{85B5E5}1.44\%} &
  \multicolumn{1}{c|}{\cellcolor[HTML]{E5F9E8}-0.17\%} &
  \multicolumn{1}{c|}{\cellcolor[HTML]{F2AF8B}84.45\%} &
  \multicolumn{1}{c|}{\cellcolor[HTML]{DC7AD2}9.04\%} &
  \cellcolor[HTML]{9AF5FA}0.75\% \\ \hline
\multicolumn{1}{|l|}{38\% Minority   with Synthetic} &
  \multicolumn{1}{c|}{\cellcolor[HTML]{91BCE8}1.28\%} &
  \multicolumn{1}{c|}{\cellcolor[HTML]{D9F6DD}0.01\%} &
  \multicolumn{1}{c|}{\cellcolor[HTML]{FCEBE3}1.77\%} &
  \multicolumn{1}{c|}{\cellcolor[HTML]{FFFFFF}0.06\%} &
  \cellcolor[HTML]{FFFFFF}-0.59\% \\ \hline
\multicolumn{1}{|l|}{40\% Minority   with Synthetic} &
  \multicolumn{1}{c|}{\cellcolor[HTML]{90BCE8}1.29\%} &
  \multicolumn{1}{c|}{\cellcolor[HTML]{A6EAAF}0.82\%} &
  \multicolumn{1}{c|}{\cellcolor[HTML]{FCEAE0}4.23\%} &
  \multicolumn{1}{c|}{\cellcolor[HTML]{FAE9F8}1.49\%} &
  \cellcolor[HTML]{6FF1F8}1.31\% \\ \hline
\multicolumn{1}{|l|}{42\% Minority   with Synthetic} &
  \multicolumn{1}{c|}{\cellcolor[HTML]{EDF4FB}0.03\%} &
  \multicolumn{1}{c|}{\cellcolor[HTML]{CFF4D4}0.17\%} &
  \multicolumn{1}{c|}{\cellcolor[HTML]{FCEDE5}-0.08\%} &
  \multicolumn{1}{c|}{\cellcolor[HTML]{FDF5FC}0.74\%} &
  \cellcolor[HTML]{ACF7FB}0.51\% \\ \hline
\multicolumn{1}{|l|}{45\% Minority   with Synthetic} &
  \multicolumn{1}{c|}{\cellcolor[HTML]{BBD6F1}0.70\%} &
  \multicolumn{1}{c|}{\cellcolor[HTML]{FFFFFF}-0.58\%} &
  \multicolumn{1}{c|}{\cellcolor[HTML]{FCECE3}1.10\%} &
  \multicolumn{1}{c|}{\cellcolor[HTML]{FEF8FD}0.48\%} &
  \cellcolor[HTML]{E3FDFE}-0.22\% \\ \hline
\multicolumn{1}{|l|}{48\% Minority   with Synthetic} &
  \multicolumn{1}{c|}{\cellcolor[HTML]{EBF3FB}0.05\%} &
  \multicolumn{1}{c|}{\cellcolor[HTML]{AFECB7}0.68\%} &
  \multicolumn{1}{c|}{\cellcolor[HTML]{FCEBE2}2.06\%} &
  \multicolumn{1}{c|}{\cellcolor[HTML]{FBEDF9}1.25\%} &
  \cellcolor[HTML]{71F1F8}1.29\% \\ \hline
\multicolumn{1}{|l|}{50\% Minority   with Synthetic} &
  \multicolumn{1}{c|}{\cellcolor[HTML]{4D93D9}2.20\%} &
  \multicolumn{1}{c|}{\cellcolor[HTML]{47D359}2.30\%} &
  \multicolumn{1}{c|}{\cellcolor[HTML]{FCEAE0}3.92\%} &
  \multicolumn{1}{c|}{\cellcolor[HTML]{FCF4FC}0.77\%} &
  \cellcolor[HTML]{9DF5FA}0.71\% \\ \hline
\end{tabular}%
}
\caption{Table showing \%improvement of classification statistics across 3 Models RF,KNN,NN}
\label{Improvemnt Table}
\end{table}

\section{Inferences from Simulation} \label{Inferences}
In the process of creating the variant Quantum-SMOTEV2 algorithm and inclusion of the feature of Outlier boosting, We have reached various findings that we want to highlight in the following observations.

\begin{itemize}
\item The Quantum-SMOTEV2 algorithm retains all the features and benefits of the Quantum-SMOTE \cite{Mohanty24_Q-SMOTE} method but removes the overhead of clustering the dataset. 

\item The proposed algorithm introduces the concept of angular distribution of data around the data centroid which can be an evolving research area for future algorithms to explore.

\item The angular distribution produces angular outliers, which are used by the algorithm to implement Angular Outlier Boosting that enhances the Quantum-SMOTEV2 algorithm to classify edge cases better and improve the classification characteristics of the model.

\item The algorithm preserves the hyperparameters from the previous version, allowing users to control many aspects of synthetic data generation, including rotation angle, minority percentage, and splitting factor. Additionally, it adds the hyperparameter Bins, which aids in binning the Angular Outliers for outlier enhancement..

\item By opting for a smaller angle of rotation,the synthetic data points are positioned in proximity to the original minority data point, hence augmenting the density of minority data points in a sparsely inhabited region.

\item By selecting a wider angle of rotation for outliers, newly created synthetic data points avoid duplication with the main algorithm.

\item The method continues to use rotation circuits for minority data points, which do not promote the utilisation of entanglement processes or analogous gates such as CNOT or ZZ, since they would have undesirable effects on rotation and lead to unforeseen results.

\item The proposed algorithm still uses a compact swap test approach where more columns can be stored in fewer qubits. 

\item The algorithm's use of low-depth circuits renders it less vulnerable to complications related to extended circuits, such as noise and decoherence. It successfully demonstrates how quantum approaches may improve conventional machine-learning methods.

\item The testing of Quantum-SMOTEV2 in three different classes of algorithms, Random Forest(Ensemble Learning), KNN(lazy learning) , and NN, proved the utility of the algorithm in different scenarios and hence established its wider applicability.

\item Application of Angular Outlier boosting after Quantum-SMOTEV2 proved marked improvement in ROC, PR, and F1 scores across all models and established the wider applicability of the procedure.

\end{itemize}

\section{Conclusion}\label{Conclusion}
The proposed variant of Quantum-SMOTE works well in highly imbalanced datasets. The resulting testing of algorithm depicts a tremendous increase in the performance of the three tested classifiers, namely NN, KNN, and RF. The fact that the increased percentage of Quantum-SMOTEV2 gives substantial gains in key performance metrics, particularly in F1 score, PR AUC, and ROC AUC, makes these metrics very important in cases of imbalanced data. Among them, Quantum-SMOTEV2 especially favors NN and RF, where the latter two have been consistently improving in accuracy, class differentiation in ROC AUC, and handling minority classes in F1 score and PR AUC. KNN has a tendency to exhibit mild improvements but is still behind when compared to others.

Outlier Boosting reinforces the strength of Quantum-SMOTEV2 by fine-tuning the model to handle edge cases and other hard-to-classify instances, which includes those from minority classes as well. The boosting of outlier instances by Outlier Boosting is complementary to the handling of synthetic data generated by Quantum-SMOTEV2 for better balancing the classification. This is most striking in RF, where the boosted outliers elevate the F1 score and PR AUC to the highest level compared to other models, denoting better precision and recall. Similarly, in the case of NN, PR AUC and ROC AUC show a great improvement while classifying minority classes drastically. Outlier Boosting becomes important in model performance optimization for models trained with Quantum-SMOTEV2, especially when there is much imbalance in the dataset. This improves the capacity of the models to correctly identify instances of the minority class without any reduction in overall model accuracy.

Overall we can conclude the Proposed Quantum-SMOTEV2 along with Angular Outlier boosting is a remarkebaly efficient aligorithm showcasing innovative use of quantum computing principles in enhancing classical machinelearing algorithms with wide variety of use cases.

\section*{Acknowledgment}
The authors express gratitude to the IBM Quantum Experience platform and its team for creating the Qiskit platform and granting free access to their simulators for executing quantum circuits and conducting the experiments detailed below. The authors express appreciation for the Centre for Quantum Software and Information (CQSI) and the Sydney Quantum Academy.

\section{Statements and Declarations}
\textbf{Competing Interests}: The authors have no financial or non-financial competing interests.\\
\textbf{Authors' contributions}:
The authors confirm their contribution to the paper as follows: 
Study conception and design: N.M., B.K.B., C.F.;\\ 
Data collection: N.M.;\\ 
Analysis and interpretation of results: N.M., B.K.B., C.F.;\\ 
Draft manuscript preparation: N.M., B.K.B., C.F., ;\\
All authors reviewed the results and approved the final version of the manuscript.\\
\textbf{Funding}: Authors declare that there has been no external funding.\\
\textbf{Availability of data and materials}: All the data provided in this manuscript is generated during the simulation and can be provided upon reasonable request.

\bibliography{sn-bibliography}
\section{Supplementary Figures } \label{Model Performance}
\subsection{Confusion matrix}
This section covers the normalized confusion matrices of three tested classifiers: RF, KNN, and NN. Confusion matrices are organised into two sections for each algorithm the first covers confusion matrices without Quantum-SMOTE and confusion matrices post application of Quantum-SMOTEV2 at 34\%, 42\% and 50\%. The second section covers confusion matrices with outlier boosting.
\begin{figure}[H]
\centering
\begin{subfigure}{0.5\linewidth}
\includegraphics[width=\linewidth]{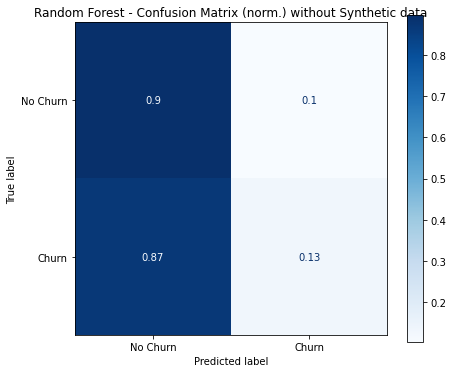} 
\caption{}
\label{qsmote_Fig10a}
\end{subfigure}\hfill
\begin{subfigure}{0.5\linewidth}
\includegraphics[width=\linewidth]{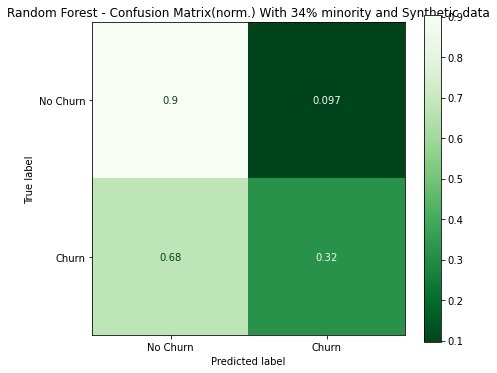} 
\caption{}
\label{qsmote_Fig10b}
\end{subfigure}\hfill
\begin{subfigure}{0.5\linewidth}
\includegraphics[width=\linewidth]{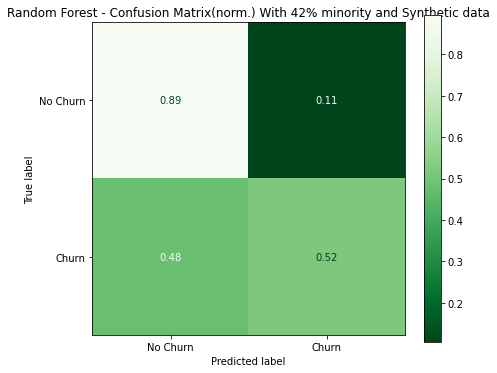} 
\caption{}
\label{qsmote_Fig10c}
\end{subfigure}\hfill
\begin{subfigure}{0.5\linewidth}
\includegraphics[width=\linewidth]{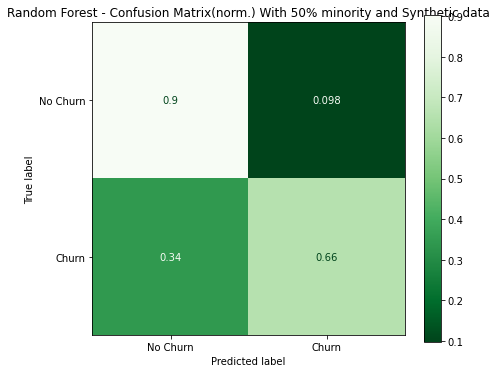} 
\caption{}
\label{qsmote_Fig10d}
\end{subfigure}\hfill
\caption{Plot illustrating Model Charts for RF model  Normalised Confusion matrices. (a) Confusion Matrix  without smote, (b) Confusion Matrix  with 34\% Q-SMOTE, (c) Confusion Matrix  with 42\% Q-SMOTE, (d) Confusion Matrix  with 50\% Q-SMOTE.}
\label{qsmote_Fig11}
\end{figure}

\begin{figure}[H]
\centering
\begin{subfigure}{0.5\linewidth}
\includegraphics[width=\linewidth]{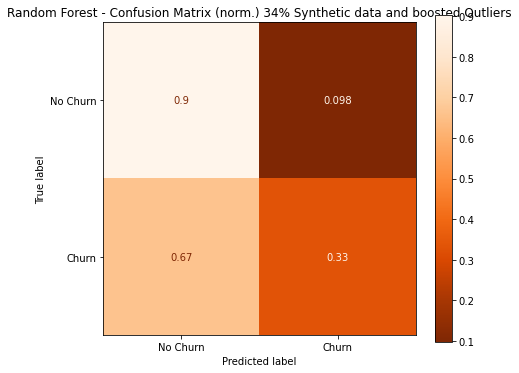} 
\caption{}
\label{qsmote_Fig10b}
\end{subfigure}\hfill
\begin{subfigure}{0.5\linewidth}
\includegraphics[width=\linewidth]{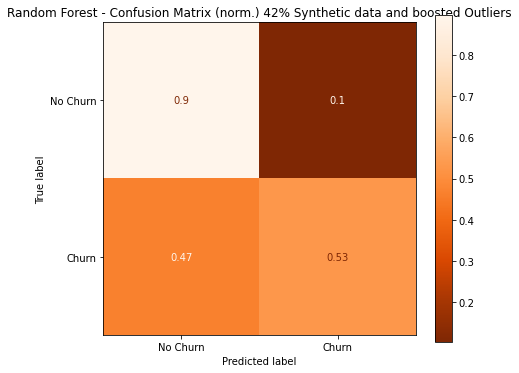} 
\caption{}
\label{qsmote_Fig10c}
\end{subfigure}\hfill
\begin{subfigure}{0.5\linewidth}
\includegraphics[width=\linewidth]{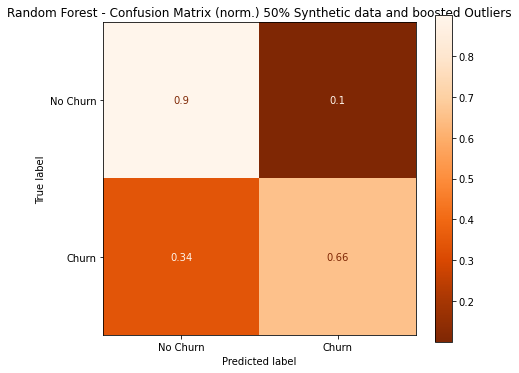} 
\caption{}
\label{qsmote_Fig10d}
\end{subfigure}\hfill
\caption{Model Charts for RF model  Normalised Confusion matrices with Oultlier Boosting. (a) Confusion Matrix with 34\% Q-SMOTEOL, (c) Confusion Matrix with 42\% Q-SMOTEOL, (d) Confusion Matrix with 50\% Q-SMOTEOL.}
\label{qsmote_Fig12}
\end{figure}

\begin{figure}[H]
\centering
\begin{subfigure}{0.5\linewidth}
\includegraphics[width=\linewidth]{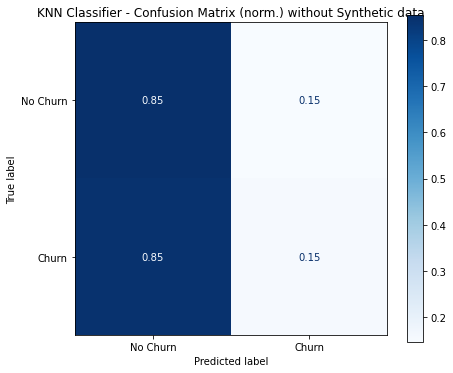} 
\caption{}
\label{qsmote_Fig10a}
\end{subfigure}\hfill
\begin{subfigure}{0.5\linewidth}
\includegraphics[width=\linewidth]{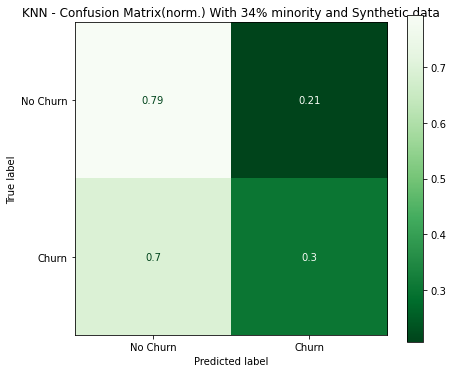} 
\caption{}
\label{qsmote_Fig10b}
\end{subfigure}\hfill
\begin{subfigure}{0.5\linewidth}
\includegraphics[width=\linewidth]{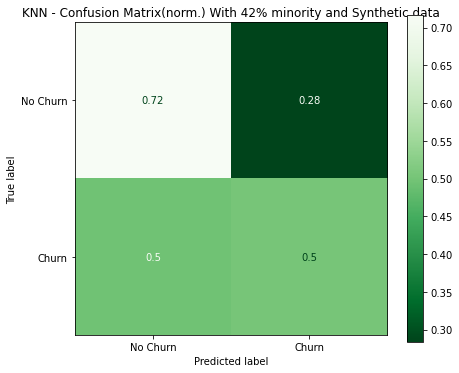} 
\caption{}
\label{qsmote_Fig10c}
\end{subfigure}\hfill
\begin{subfigure}{0.5\linewidth}
\includegraphics[width=\linewidth]{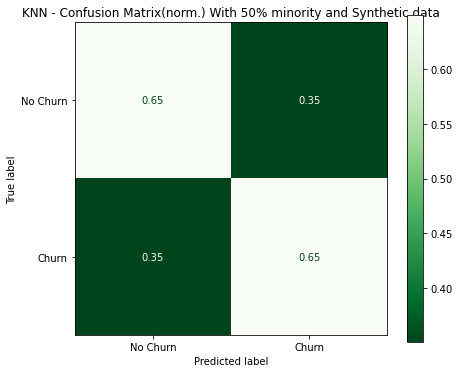} 
\caption{}
\label{qsmote_Fig10d}
\end{subfigure}\hfill
\caption{Model Charts for KNN Clssification  Normalised Confusion matrices. (a) Confusion Matrix  without smote, (b) Confusion Matrix  with 34\% Q-SMOTE, (c) Confusion Matrix  with 42\% Q-SMOTE, (d) Confusion Matrix  with 50\% Q-SMOTE.}
\label{qsmote_Fig13}
\end{figure}

\begin{figure}[H]
\centering
\begin{subfigure}{0.5\linewidth}
\includegraphics[width=\linewidth]{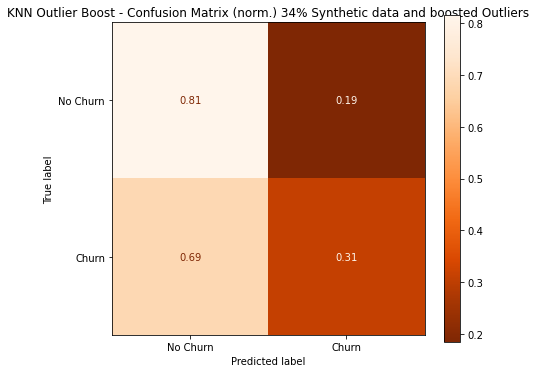} 
\caption{}
\label{qsmote_Fig10b}
\end{subfigure}\hfill
\begin{subfigure}{0.5\linewidth}
\includegraphics[width=\linewidth]{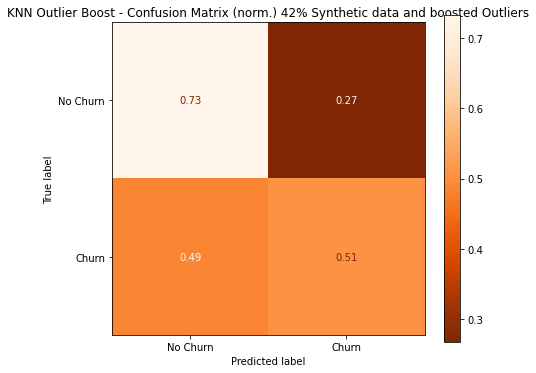} 
\caption{}
\label{qsmote_Fig10c}
\end{subfigure}\hfill
\begin{subfigure}{0.5\linewidth}
\includegraphics[width=\linewidth]{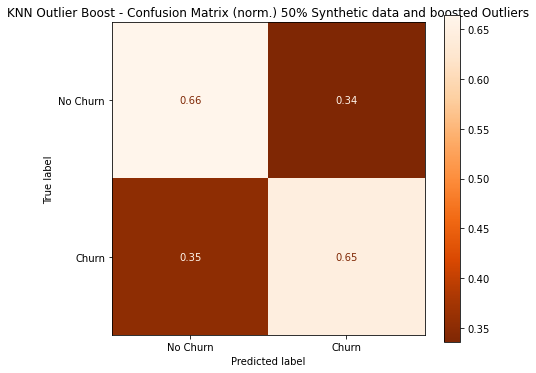} 
\caption{}
\label{qsmote_Fig10d}
\end{subfigure}\hfill
\caption{Model Charts for KNN Clssification  Normalised Confusion matrices with Oultlier Boosting. (a) Confusion Matrix with 34\% Q-SMOTEOL, (c) Confusion Matrix with 42\% Q-SMOTEOL, (d) Confusion Matrix with 50\% Q-SMOTEOL.}
\label{qsmote_Fig14}
\end{figure}

\begin{figure}[H]
\centering
\begin{subfigure}{0.5\linewidth}
\includegraphics[width=\linewidth]{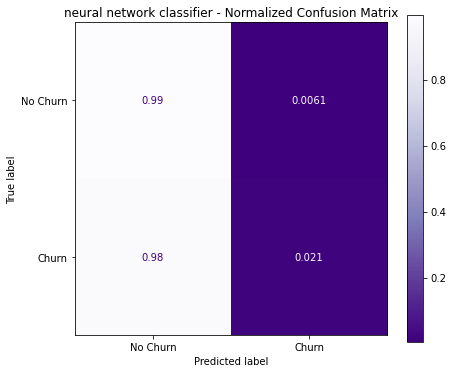} 
\caption{}
\label{qsmote_Fig10a}
\end{subfigure}\hfill
\begin{subfigure}{0.5\linewidth}
\includegraphics[width=\linewidth]{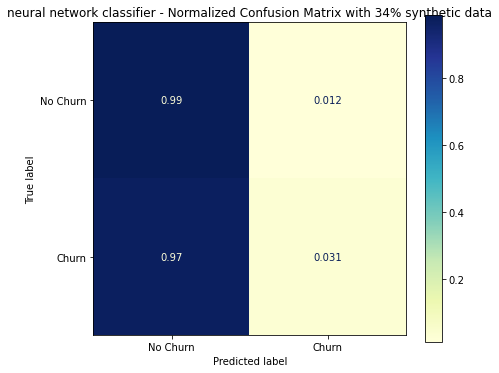} 
\caption{}
\label{qsmote_Fig10b}
\end{subfigure}\hfill
\begin{subfigure}{0.5\linewidth}
\includegraphics[width=\linewidth]{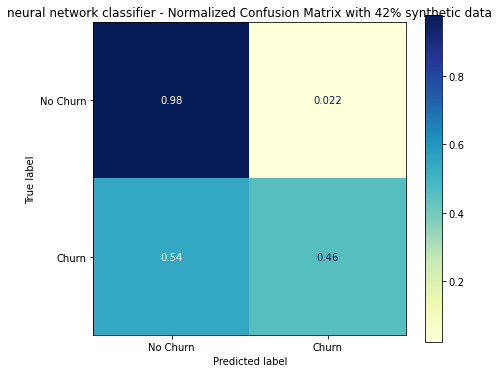} 
\caption{}
\label{qsmote_Fig10c}
\end{subfigure}\hfill
\begin{subfigure}{0.5\linewidth}
\includegraphics[width=\linewidth]{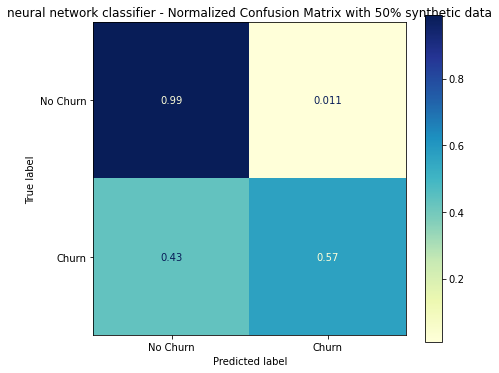} 
\caption{}
\label{qsmote_Fig10d}
\end{subfigure}\hfill
\caption{Model Charts for NN  Normalised Confusion matrices. (a) Confusion Matrix  without smote, (b) Confusion Matrix  with 34\% Q-SMOTE, (c) Confusion Matrix  with 42\% Q-SMOTE, (d) Confusion Matrix  with 50\% Q-SMOTE.}
\label{qsmote_Fig15}
\end{figure}

\begin{figure}[H]
\centering
\begin{subfigure}{0.5\linewidth}
\includegraphics[width=\linewidth]{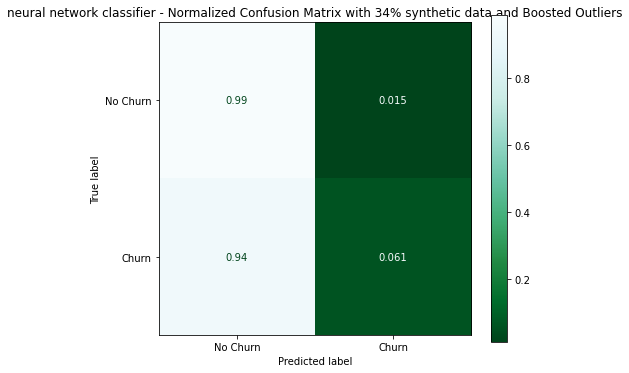} 
\caption{}
\label{qsmote_Fig10b}
\end{subfigure}\hfill
\begin{subfigure}{0.5\linewidth}
\includegraphics[width=\linewidth]{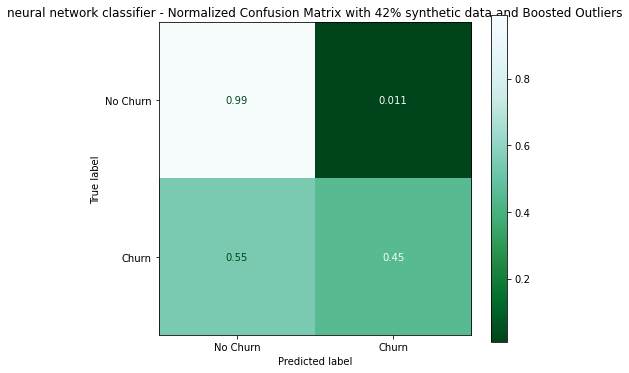} 
\caption{}
\label{qsmote_Fig10c}
\end{subfigure}\hfill
\begin{subfigure}{0.5\linewidth}
\includegraphics[width=\linewidth]{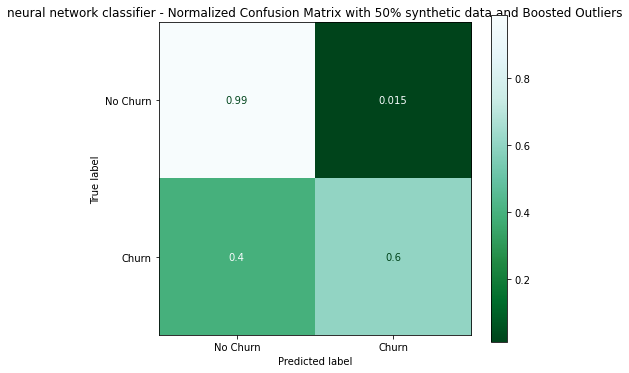} 
\caption{}
\label{qsmote_Fig10d}
\end{subfigure}\hfill
\caption{Model Charts for NN  Normalised Confusion matrices with Oultlier Boosting. (a) Confusion Matrix with 34\% Q-SMOTEOL, (c) Confusion Matrix with 42\% Q-SMOTEOL, (d) Confusion Matrix with 50\% Q-SMOTEOL.}
\label{qsmote_Fig16}
\end{figure}

\subsection{ROC}
This section covers the ROC-AUC characterstics of three tested classifiers: RF, KNN, and NN. ROC-AUC characteristics are organized into two sections for each algorithm. The first covers ROC-AUC  without Quantum-SMOTE and confusion matrices post application of Quantum-SMOTEV2 at 34\%, 42\%, and 50\%. The second section covers ROC-AUC with outlier boosting.

\begin{figure}[H]
\centering
\begin{subfigure}{0.5\linewidth}
\includegraphics[width=\linewidth]{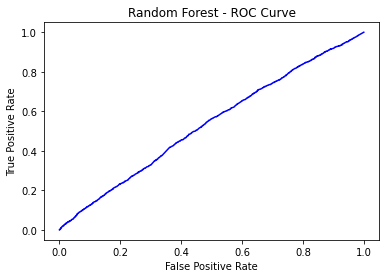} 
\caption{}
\label{qsmote_Fig18a}
\end{subfigure}\hfill
\begin{subfigure}{0.5\linewidth}
\includegraphics[width=\linewidth]{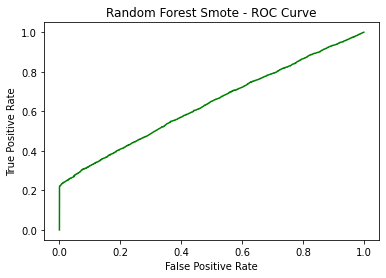} 
\caption{}
\label{qsmote_Fig18b}
\end{subfigure}\hfill
\begin{subfigure}{0.5\linewidth}
\includegraphics[width=\linewidth]{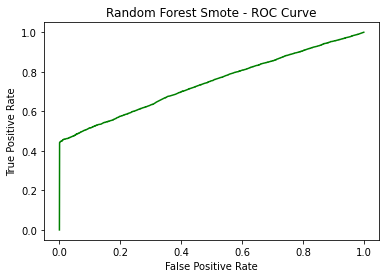} 
\caption{}
\label{qsmote_Fig18c}
\end{subfigure}\hfill
\begin{subfigure}{0.5\linewidth}
\includegraphics[width=\linewidth]{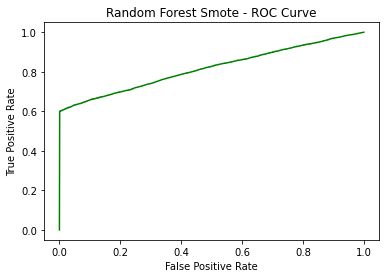} 
\caption{}
\label{qsmote_Fig18d}
\end{subfigure}\hfill
\caption{AUC-ROC for RF model with/without smote for comparison. (a) AUC-ROC  without smote, (b) AUC-ROC with smote and 34\% Minority, (c) AUC-ROC  with smote and 42\% Minority, (d) AUC-ROC  with smote and 50\% Minority.}
\label{qsmote_Fig17}
\end{figure}

\begin{figure}[H]
\centering
\begin{subfigure}{0.5\linewidth}
\includegraphics[width=\linewidth]{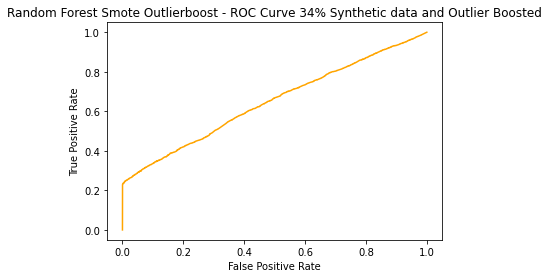} 
\caption{}
\label{qsmote_Fig18b}
\end{subfigure}\hfill
\begin{subfigure}{0.5\linewidth}
\includegraphics[width=\linewidth]{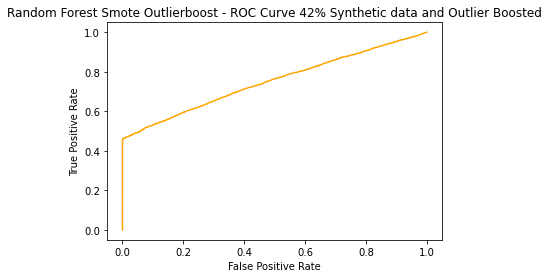} 
\caption{}
\label{qsmote_Fig18c}
\end{subfigure}\hfill
\begin{subfigure}{0.5\linewidth}
\includegraphics[width=\linewidth]{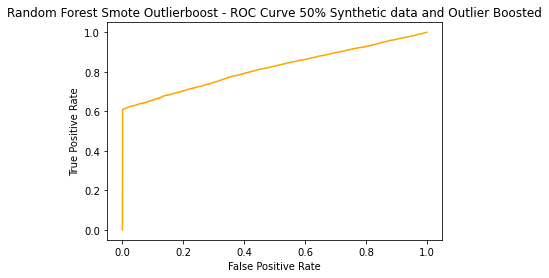} 
\caption{}
\label{qsmote_Fig18d}
\end{subfigure}\hfill
\caption{AUC-ROC for RF model with smote and Outlier boosting for comparison. (a)  AUC-ROC with smote and 34\% Minority, (c) AUC-ROC  with smote and 42\% Minority, (d) AUC-ROC  with smote and 50\% Minority.}
\label{qsmote_Fig18}
\end{figure}

\begin{figure}[H]
\centering
\begin{subfigure}{0.5\linewidth}
\includegraphics[width=\linewidth]{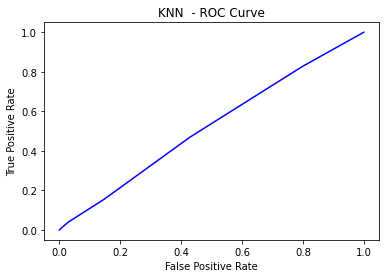} 
\caption{}
\label{qsmote_Fig18a}
\end{subfigure}\hfill
\begin{subfigure}{0.5\linewidth}
\includegraphics[width=\linewidth]{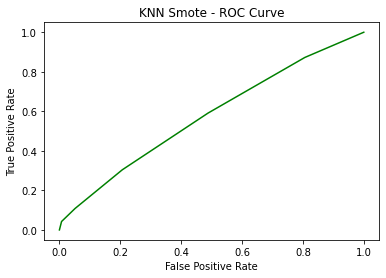} 
\caption{}
\label{qsmote_Fig18b}
\end{subfigure}\hfill
\begin{subfigure}{0.5\linewidth}
\includegraphics[width=\linewidth]{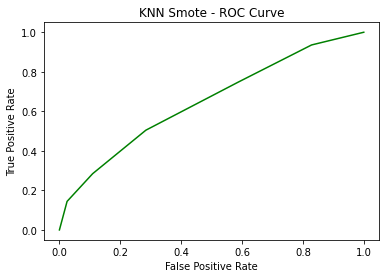} 
\caption{}
\label{qsmote_Fig18c}
\end{subfigure}\hfill
\begin{subfigure}{0.5\linewidth}
\includegraphics[width=\linewidth]{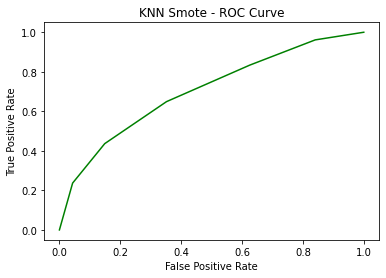} 
\caption{}
\label{qsmote_Fig18d}
\end{subfigure}\hfill
\caption{AUC-ROC for KNN Classification with/without smote for comparison. (a) AUC-ROC  without smote, (b) AUC-ROC with smote and 34\% Minority, (c) AUC-ROC  with smote and 42\% Minority, (d) AUC-ROC  with smote and 50\% Minority.}
\label{qsmote_Fig19}
\end{figure}

\begin{figure}[H]
\centering
\begin{subfigure}{0.5\linewidth}
\includegraphics[width=\linewidth]{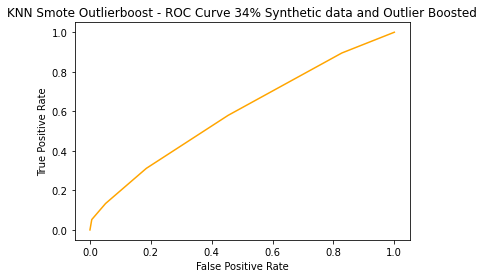} 
\caption{}
\label{qsmote_Fig18b}
\end{subfigure}\hfill
\begin{subfigure}{0.5\linewidth}
\includegraphics[width=\linewidth]{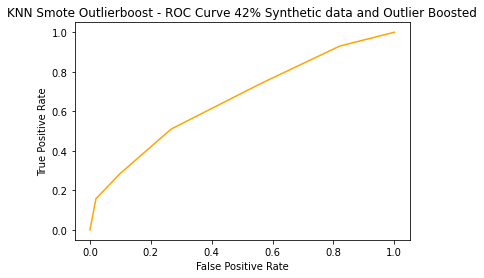} 
\caption{}
\label{qsmote_Fig18c}
\end{subfigure}\hfill
\begin{subfigure}{0.5\linewidth}
\includegraphics[width=\linewidth]{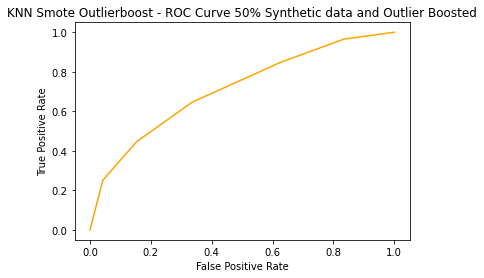} 
\caption{}
\label{qsmote_Fig18d}
\end{subfigure}\hfill
\caption{AUC-ROC for KNN Classification with smote and Outlier boosting for comparison. (a)  AUC-ROC with smote and 34\% Minority, (c) AUC-ROC  with smote and 42\% Minority, (d) AUC-ROC  with smote and 50\% Minority.}
\label{qsmote_Fig20}
\end{figure}

\begin{figure}[H]
\centering
\begin{subfigure}{0.5\linewidth}
\includegraphics[width=\linewidth]{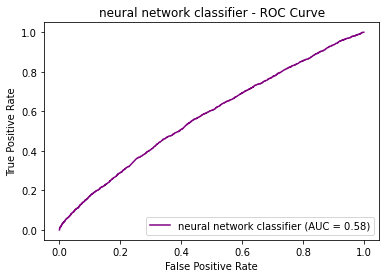} 
\caption{}
\label{qsmote_Fig18a}
\end{subfigure}\hfill
\begin{subfigure}{0.5\linewidth}
\includegraphics[width=\linewidth]{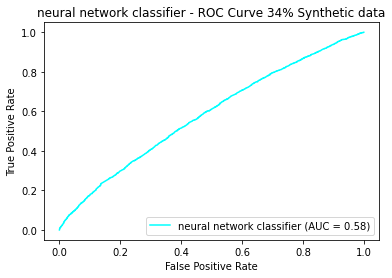} 
\caption{}
\label{qsmote_Fig18b}
\end{subfigure}\hfill
\begin{subfigure}{0.5\linewidth}
\includegraphics[width=\linewidth]{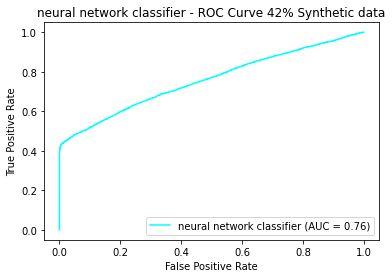} 
\caption{}
\label{qsmote_Fig18c}
\end{subfigure}\hfill
\begin{subfigure}{0.5\linewidth}
\includegraphics[width=\linewidth]{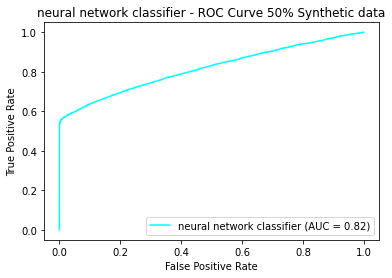} 
\caption{}
\label{qsmote_Fig18d}
\end{subfigure}\hfill
\caption{AUC-ROC for NN with/without smote for comparison. (a) AUC-ROC  without smote, (b) AUC-ROC with smote and 34\% Minority, (c) AUC-ROC  with smote and 42\% Minority, (d) AUC-ROC  with smote and 50\% Minority.}
\label{qsmote_Fig21}
\end{figure}

\begin{figure}[H]
\centering
\begin{subfigure}{0.5\linewidth}
\includegraphics[width=\linewidth]{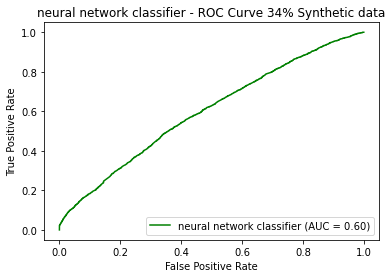} 
\caption{}
\label{qsmote_Fig18b}
\end{subfigure}\hfill
\begin{subfigure}{0.5\linewidth}
\includegraphics[width=\linewidth]{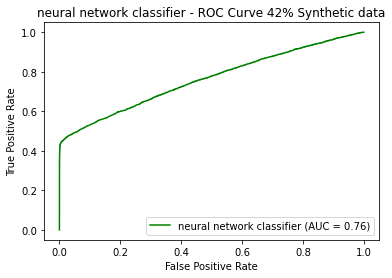} 
\caption{}
\label{qsmote_Fig18c}
\end{subfigure}\hfill
\begin{subfigure}{0.5\linewidth}
\includegraphics[width=\linewidth]{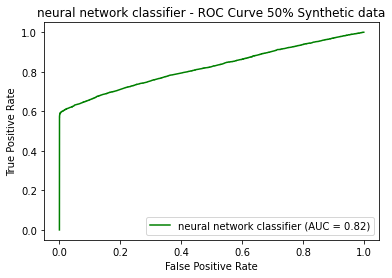} 
\caption{}
\label{qsmote_Fig18d}
\end{subfigure}\hfill
\caption{AUC-ROC for NN with smote and Outlier boosting for comparison. (a)  AUC-ROC with smote and 34\% Minority, (c) AUC-ROC  with smote and 42\% Minority, (d) AUC-ROC  with smote and 50\% Minority.}
\label{qsmote_Fig22}
\end{figure}

\subsection{Precision-Recall}

This section covers the precision-recall  PR-AUC characteristics of three tested classifiers: RF, KNN, and NN. PR-AUC characteristics are organized into two sections for each algorithm. The first covers PR-AUC  without Quantum-SMOTE and confusion matrices post application of Quantum-SMOTEV2 smote at 34\%, 42\%, and 50\%. The second section covers PR-AUC with outlier boosting.

\begin{figure}[H]
\centering
\begin{subfigure}{0.5\linewidth}
\includegraphics[width=\linewidth]{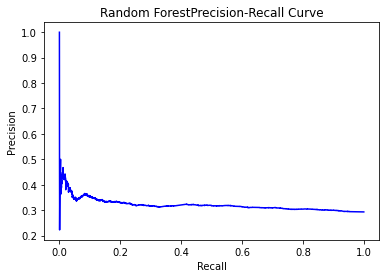} 
\caption{}
\label{qsmote_Fig15a}
\end{subfigure}\hfill
\begin{subfigure}{0.5\linewidth}
\includegraphics[width=\linewidth]{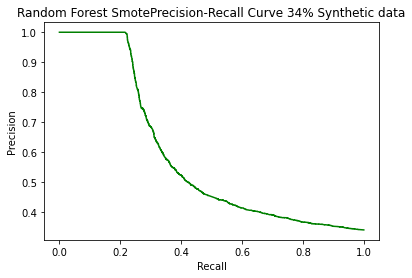} 
\caption{}
\label{qsmote_Fig15b}
\end{subfigure}\hfill
\begin{subfigure}{0.5\linewidth}
\includegraphics[width=\linewidth]{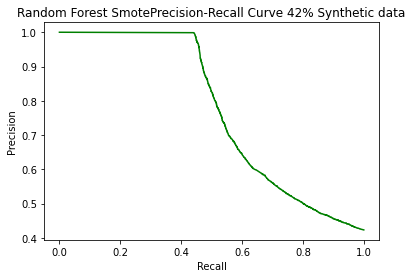} 
\caption{}
\label{qsmote_Fig15c}
\end{subfigure}\hfill
\begin{subfigure}{0.5\linewidth}
\includegraphics[width=\linewidth]{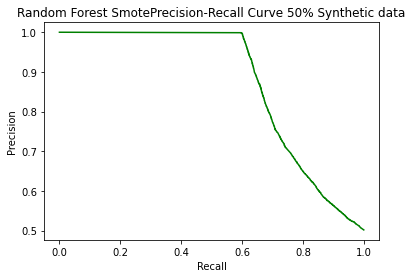} 
\caption{}
\label{qsmote_Fig15d}
\end{subfigure}\hfill
\caption{PR-AUC for RF model with/without smote for comparison. (a) PR-AUC without smote, (b) PR-AUC  with smote and 34\% Minority, (c) PR-AUC  with smote and 42\% Minority, (d) PR-AUC with smote and 50\% Minority.}
\label{qsmote_Fig23}
\end{figure}

\begin{figure}[H]
\centering
\begin{subfigure}{0.5\linewidth}
\includegraphics[width=\linewidth]{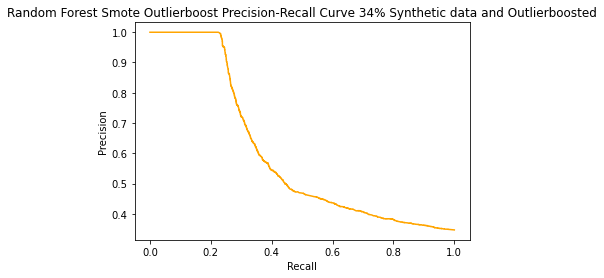} 
\caption{}
\label{qsmote_Fig15b}
\end{subfigure}\hfill
\begin{subfigure}{0.5\linewidth}
\includegraphics[width=\linewidth]{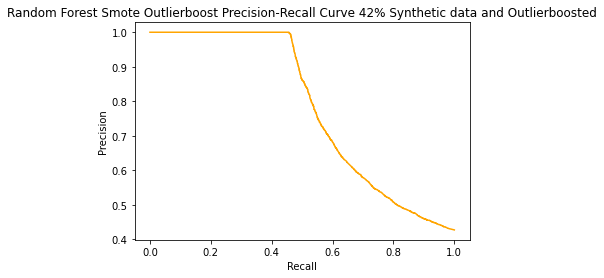} 
\caption{}
\label{qsmote_Fig15c}
\end{subfigure}\hfill
\begin{subfigure}{0.5\linewidth}
\includegraphics[width=\linewidth]{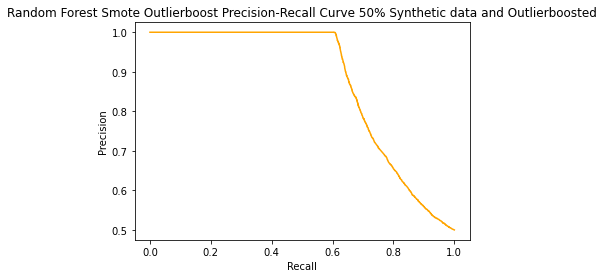} 
\caption{}
\label{qsmote_Fig15d}
\end{subfigure}\hfill
\caption{PR-AUC for RF model with smote and outlierboost. (a)  PR-AUC  with Q-SMOTEOL and 34\% Minority, (c) PR-AUC  with Q-SMOTEOL and 42\% Minority, (d) PR-AUC with Q-SMOTEOL and 50\% Minority.}
\label{qsmote_Fig24}
\end{figure}

\begin{figure}[H]
\centering
\begin{subfigure}{0.5\linewidth}
\includegraphics[width=\linewidth]{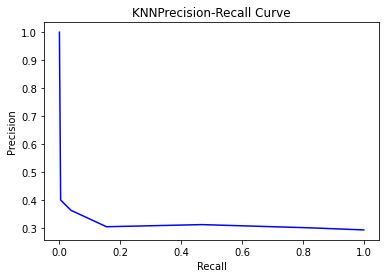} 
\caption{}
\label{qsmote_Fig15a}
\end{subfigure}\hfill
\begin{subfigure}{0.5\linewidth}
\includegraphics[width=\linewidth]{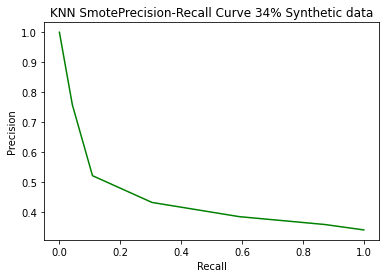} 
\caption{}
\label{qsmote_Fig15b}
\end{subfigure}\hfill
\begin{subfigure}{0.5\linewidth}
\includegraphics[width=\linewidth]{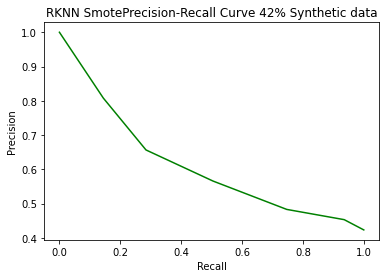} 
\caption{}
\label{qsmote_Fig15c}
\end{subfigure}\hfill
\begin{subfigure}{0.5\linewidth}
\includegraphics[width=\linewidth]{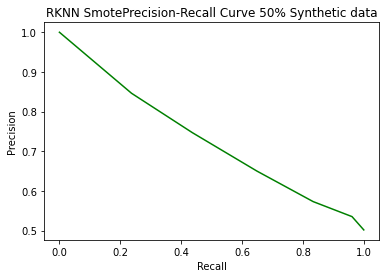} 
\caption{}
\label{qsmote_Fig15d}
\end{subfigure}\hfill
\caption{PR-AUC for KNN Classifier with/without smote for comparison. (a) PR-AUC without smote, (b) PR-AUC  with smote and 34\% Minority, (c) PR-AUC with smote and 42\% Minority, (d) PR-AUC with smote and 50\% Minority.}
\label{qsmote_Fig25}
\end{figure}

\begin{figure}[H]
\centering
\begin{subfigure}{0.5\linewidth}
\includegraphics[width=\linewidth]{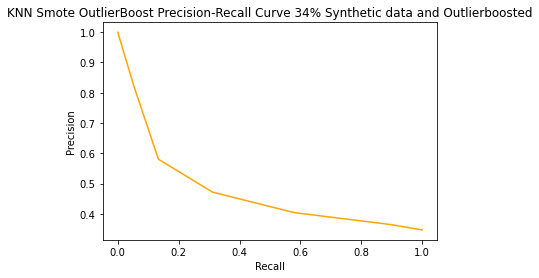} 
\caption{}
\label{qsmote_Fig15b}
\end{subfigure}\hfill
\begin{subfigure}{0.5\linewidth}
\includegraphics[width=\linewidth]{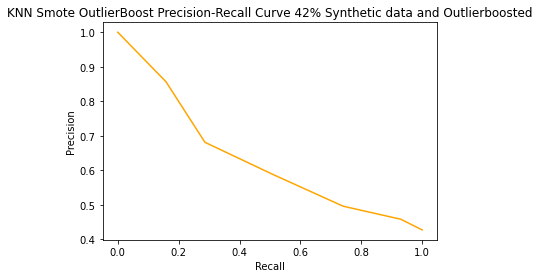} 
\caption{}
\label{qsmote_Fig15c}
\end{subfigure}\hfill
\begin{subfigure}{0.5\linewidth}
\includegraphics[width=\linewidth]{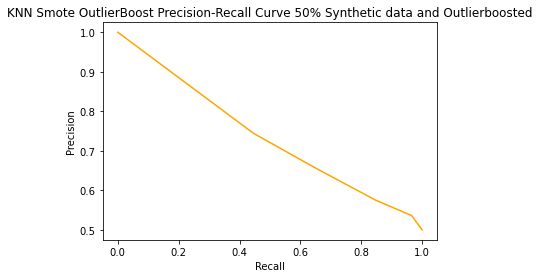} 
\caption{}
\label{qsmote_Fig15d}
\end{subfigure}\hfill
\caption{PR-AUC for KNN Classifier with smote and outlier boost. (a)  PR-AUC  with Q-SMOTEOL and 34\% Minority, (c) PR-AUC  with Q-SMOTEOL and 42\% Minority, (d) PR-AUC with Q-SMOTEOL and 50\% Minority.}
\label{qsmote_Fig26}
\end{figure}

\begin{figure}[H]
\centering
\begin{subfigure}{0.5\linewidth}
\includegraphics[width=\linewidth]{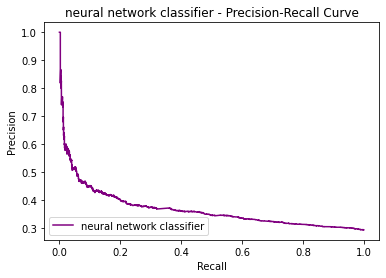} 
\caption{}
\label{qsmote_Fig15a}
\end{subfigure}\hfill
\begin{subfigure}{0.5\linewidth}
\includegraphics[width=\linewidth]{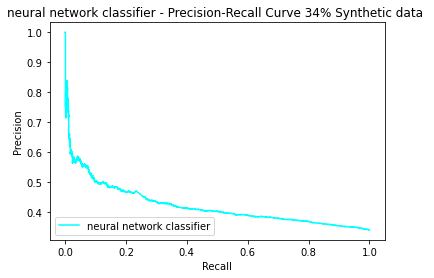} 
\caption{}
\label{qsmote_Fig15b}
\end{subfigure}\hfill
\begin{subfigure}{0.5\linewidth}
\includegraphics[width=\linewidth]{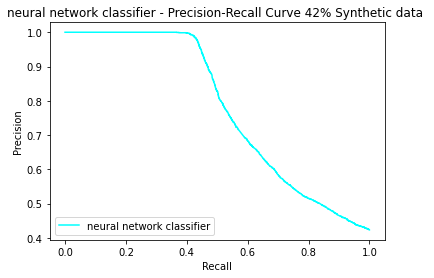} 
\caption{}
\label{qsmote_Fig15c}
\end{subfigure}\hfill
\begin{subfigure}{0.5\linewidth}
\includegraphics[width=\linewidth]{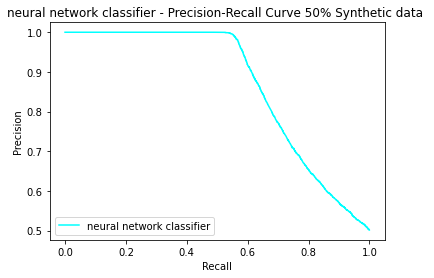} 
\caption{}
\label{qsmote_Fig15d}
\end{subfigure}\hfill
\caption{Plot illustrating Precision-Recall Curve (AUC) for NN with/without smote for comparison. (a) PR-AUC without smote, (b) PR-AUC  with smote and 34\% Minority, (c) PR-AUC  with smote and 42\% Minority, (d) PR-AUC with smote and 50\% Minority.}
\label{qsmote_Fig27}
\end{figure}

\begin{figure}[H]
\centering
\begin{subfigure}{0.5\linewidth}
\includegraphics[width=\linewidth]{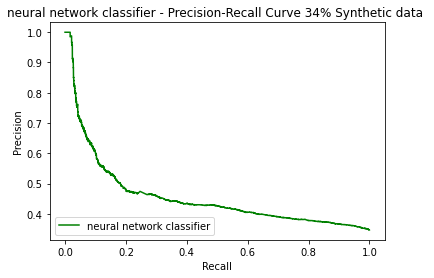} 
\caption{}
\label{qsmote_Fig15b}
\end{subfigure}\hfill
\begin{subfigure}{0.5\linewidth}
\includegraphics[width=\linewidth]{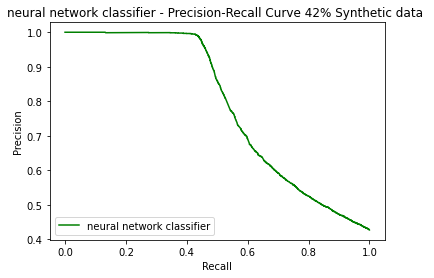} 
\caption{}
\label{qsmote_Fig15c}
\end{subfigure}\hfill
\begin{subfigure}{0.5\linewidth}
\includegraphics[width=\linewidth]{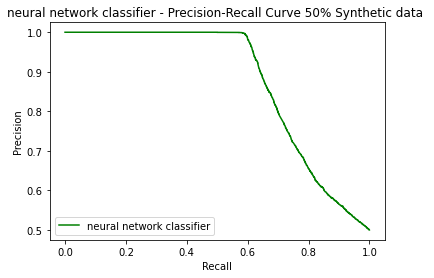} 
\caption{}
\label{qsmote_Fig15d}
\end{subfigure}\hfill
\caption{PR-AUC for NN with smote and outlierboost. (a)  PR-AUC  with Q-SMOTEOL and 34\% Minority, (c) PR-AUC  with Q-SMOTEOL and 42\% Minority, (d) PR-AUC with Q-SMOTEOL and 50\% Minority.}
\label{qsmote_Fig28}
\end{figure}

\end{document}